\documentclass[12pt,oneside,openany]{article}

\usepackage{authblk}
\title{A System of ODEs for Representing Trends of CGM Signals}

\author[1]{Giulia Elena Aliffi}

\author[1]{Giovanni Nastasi}

\author[1]{Vittorio Romano}

\author[2]{Dario Pitocco}

\author[3]{Alessandro Rizzi}

\author[4]{Elvin J. Moore}

\author[5,6,7]{Andrea De Gaetano}

\affil[1]{Dipartimento di Matematica e Informatica, Universit\'a degli Studi di Catania, Catania ITALY}
\affil[2]{Universit\'a Cattolica del Sacro Cuore, Rome ITALY}
\affil[3]{Fondazione Policlinico Universitario Agostino gemelli IRCCS, Rome ITALY}
\affil[4]{Department of Mathematics, King Mongkut's University Of Technology North Bangkok, Bangkok THAILAND }

\affil[5]{CNR-IASI Istituto di analisi dei sistemi ed informatica "Antonio Ruberti", Roma, ITALY}
\affil[6]{CNR-IRIB Istituto per la Ricerca e l'Innovazione Biomedica, Palermo ITALY }
\affil[7]{ \'Obuda University, Budapest HUNGARY}

\usepackage{amssymb}

\usepackage{amsthm}
\usepackage{authblk}

\usepackage{amsfonts}
\usepackage[T1]{fontenc}
\usepackage[utf8]{inputenc}
\usepackage{array}
\usepackage{float}
\usepackage{enumitem}
\usepackage{pdfpages}
\usepackage{lscape}
\usepackage{supertabular}
\usepackage{caption}
\usepackage{subcaption}
\usepackage{hyperref}

\date{}
\begin{document}
\maketitle

\begin{abstract}
Diabetes Mellitus is a metabolic disorder which may result in severe and potentially fatal complications if not well-treated and monitored. In this study, a quantitative analysis of the data collected using CGM (Continuous Glucose Monitoring) devices from eight subjects with type 2 diabetes in good metabolic control at the University Polyclinic Agostino Gemelli, Catholic University of the Sacred Heart, was carried out. In particular, a system of ordinary differential equations whose state variables are affected by a sequence of stochastic perturbations was proposed and used to extract more informative inferences from the patients' data. For this work, Matlab and R programs were used to find the most appropriate values of the parameters (according to the Akaike Information Criterion (AIC) and the Bayesian Information Criterion (BIC)) for each patient. Fitting was carried out by Particle Swarm Optimization to minimize the ordinary least squares error between the observed CGM data and the data from the ODE model. Goodness of fit tests were made in order to assess which probability distribution was best suitable for representing the waiting times computed from the model parameters. Finally, both parametric and non-parametric density estimation of the frequency histograms associated with the variability of the glucose elimination rate from blood were conducted and their representative parameters assessed from the data. The results show that the chosen models succeed in capturing most of the glucose fluctuations for almost every patient.
\end{abstract}

\section*{Keywords}
Diabetes Mellitus , Continuous Glucose Monitoring , Random Ordinary Differential Equations  , Particle Swarm Optimization Method, Maximum Likelihood Estimation, Akaike Information Criterion

\section{Introduction}
\label{Introduction}

\noindent
Marked by chronically high basal and post-prandial glycemia due to insulin production deficiency and$\slash$or tissue insulin resistance, Diabetes Mellitus is increasing in prevalence throughout the world, especially as a function of evolving food intake and physical exercise patterns~\cite{insight,who,epidemic}.\\

\noindent
In its various forms, Diabetes Mellitus (DM) is now the eighth major cause of death worldwide~\cite{who}. Undiagnosed or ineffectively treated DM can result in severe and potentially fatal complications (heart disease, strokes, microvascular insufficiency leading to blindness or  limb amputations, infections, even  the development of specific types of cancer and physical and cognitive disability). Among pregnant women DM can be the cause of fetal loss, congenital malformations, stillbirth and perinatal death~\cite{insight,who}. Uncontrolled hyperglycemia may lead to diabetic ketoacidosis and hyperosmolar coma, while treatment with injected insulin carries a concrete risk of hypoglycemia, with attendant seizures, coma and possibly death.\\

\noindent
According to the American Diabetes Association, the disease can be classified into three main categories: type 1 (T1DM), type 2 (T2DM) and gestational diabetes (GDM)~\cite{epidemic}.\\

\noindent
T1DM is mainly characterized by a deficiency of insulin production, probably caused by autoimmune destruction of pancreatic $\beta$-cells~\cite{insight,who,epidemic}. Type 1 diabetic patients account for approximately $5$-$10\%$ of the diabetic population. In the majority of patients (around $80$-$90\%$) the disease appears when they are children or adolescents~\cite{epidemic}. T1DM is not currently preventable or curable: for people who are affected, daily insulin injections or continuous insulin infusion via pump are essential for survival~\cite{who}. The causes of the autoimmune response, set in motion by T-lymphocytes~\cite{insight}, have not yet been precisely identified. However, several studies have suggested genetics, viral infections and environmental stress as possible risk factors~\cite{who, epidemic}. \\

\noindent
T2DM affects $90$-$95\%$ of diabetics. It is characterized by an imbalance associated with the development of insulin resistance of peripheral tissues, causing an imbalance between insulin requirements and insulin supply (production being typically increased in the early phases of the disease)~\cite{insight,epidemic}. T2DM may evolve slowly and remain sub-clinical for several years before the appearance of hyperglycemia~\cite{who}. Typically, T2DM patients do not require insulin treatment unless pancreatic $\beta$-cells loss occurs~\cite{epidemic}. Although T2DM mostly affects adults, increasing prevalence in adolescents and young adults is now becoming apparent. Obesity, sedentary lifestyle and unhealthy food choices during childhood are considered to be the major causes of
this phenomenon~\cite{epidemic}, even though genetic predisposition and the metabolic environment during gestation may also play an important role~\cite{who}. Unlike T1DM, which is not currently preventable, T2DM may be prevented or its development slowed down. Insulin sensitivity can be improved through a healthy diet and regular physical activity~\cite{who}. Very often these lifestyle changes, if maintained, are more effective than pharmacological treatments~\cite{who}.\\

\noindent
Gestational Diabetes Mellitus (GDM) is a non-permanent condition which arises in $5$ to $10\%$ of pregnant women. It consists of glycemic levels, which are above average but not high enough for the patient to be classified as outright diabetic. It is most often diagnosed by prenatal screening rather than through the appearance of symptoms. It may be influenced by maternal age, overweight, obesity, diabetes in the family, sedentary lifestyle, previous occurrence of GDM, stillbirth and polycystic ovary syndrome~\cite{who, epidemic}. Among the women involved, $40$ to $60\%$ are likely to develop T2DM in the following decade~\cite{insight,who}. In addition, GDM may lead to macrosomia, preterm births and increased frequency of cesarean deliveries~\cite{epidemic}.\\

\noindent
Among the diabetic population, it is also possible to identify a minority of about $1-2\%$ of individuals affected by Maturity Onset Diabetes of the Young (MODY). MODY is an autosomal dominant genetic disease that is associated with pancreatic $\beta$-cells dysfunction. Due to its symptomatic similarity with T1DM and T2DM, its diagnosis may be time consuming and may require specific clinical tests~\cite{mody1}. Typical characteristics of subjects suffering from this disease are strong family history of diabetes, insulin independence, lack of $\beta$-cells autoimmunity, no sign of insulin resistance mechanisms and occurrence of diabetes between the second and fifth decades. Depending on the gene involved in the mutation causing the $\beta-$cells dysfunction, the clinical picture, the prognosis the response to therapy can vary~\cite{mody1,mody2}.   \\

\noindent
Monitoring blood glucose levels plays a fundamental role in controlling the progression of the disease and the efficacy of therapy. T1DM, GDM and late-stage T2DM patients require strict glycemic control and frequent measurements of glycemia~\cite{who}. Although originally developed for T1DM patients, Continuous Glucose Monitoring (CGM) is now a common diagnostic tool for all patients receiving significant insulin therapy~\cite{horizon,cgmweb}.\\

\noindent
A CGM device provides between 288 and 1440 measurements of subcutaneous glucose concentration (nearly reflecting blood glucose concentration) per day, thus allowing the possibility of assessing glycemic variability and identifying the times of the day in which the risk of hypo- and hyper-glycemia is maximal~\cite{horizon,cgmsuccess}. As a result of monitoring, the treatment regimen (diet, insulin dosage, other medications, physical activity, etc.) can be adjusted.\\

\noindent
While visual examination of a CGM tracing already delivers important information to the trained practitioner, a quantitative assessment of the tracing would allow more precise and possibly more informative inferences to be drawn. This is the domain where mathematical and statistical analysis of CGM is required~\cite{cgmsuccess}.\\

\noindent
Extensive reviews and articles concerning the several approaches, which have been adopted to best describe the dynamics of the glucose-insulin system, have been given in the papers of Huard et al.~\cite{Huard} and Palumbo et al.~\cite{Palumbo}, where several methods to address this issue are reported. A commonly referenced model, not exempt from criticism~\cite{Palumbo, Arino, Panunzi}, is the nonlinear minimal model developed by Bergman et al.~\cite{Pacini,Bergam}. This model has been used by Boston et al.~\cite{Boston} to attempt to derive indices of glucose effectiveness and insulin sensitivity from relevant experimental data. Other authors have also developed modified versions of the model, where classical derivatives have been replaced with Caputo fractional derivatives~\cite{Cho, Coman2017} or where a glucose input disturbance has been introduced~\cite{Rooka}. \\

\noindent
Alternative models have been proposed and analysed in the works of Pompa et al.~\cite{Pompa}, Boiroux et al.~\cite{Boiroux}, Palumbo et al.~\cite{Palumbo2}, Panunzi et al.~\cite{Panunzi,Panunzi2}, Saleem et al.~\cite{Saleem} and
De Gaetano et al.~\cite{DeGaetano,DeGaetano1}. Specifically, Pompa et al.~\cite{Pompa} gave a comprehensive description and comparison between the Sorensen, Hovorka and UVAPadova models, and Boiroux et al.~\cite{Boiroux} proposed a Stochastic Differential Equation-Grey Box (SDE-GB) model reformulation of the Medtronic Virtual Patient (MPV) model developed by Kanderian et al.~\cite{Kanderian}. Further, Palumbo et al.~\cite{Palumbo2} and Panunzi et al.~\cite{Panunzi,Panunzi2} developed a single-delay model and Saleem et al.~\cite{Saleem} employed a system of Caputo-Fabrizio fractional order differential equations. Finally, De Gaetano et al.~\cite{DeGaetano,DeGaetano1} proposed a combination of Caputo fractional order and ordinary differential equations with external shocks in~\cite{DeGaetano} and an ordinary differential equations compartmental model in~\cite{DeGaetano1}.\\

\noindent

\noindent
In the present work, we generalize the deterministic first-order differential equation model proposed by Sakulrang et al.~\cite{Sakulrang1} for the analysis of CGM data from Type 1 diabetes patients. The new approach involves the replacement of the parameter $k_{XG}$ of~\cite{Sakulrang1}, which represented the rate of glucose elimination from the blood into the surrounding tissue, by a state variable $H(t)$.  This state variable is defined by a differential equation to which stochastic terms representing blood glucose fluctuations caused by unknown factors are added. \\

\noindent
The plan of the paper is as follows.  In section \ref{Materials and Methods}, we give details of the methods used for collection of data, the ODE model and the analytical and numerical methods used to solve it.  In section \ref{Results and Discussion}, we give a detailed discussion of the results. Finally, in section \ref{Conclusion}, we give conclusions.

%%%%%%%%%%%%%%%%%%%%%%%%%%%%%%%%%%%%%%%
\section{Materials and Methods}
\label{Materials and Methods}

%%%%%%
\subsection{Collection of Data}
\label{Collection of Data}

\noindent
For the analysis, CGM data from T2Dm subjects with good metabolic control were collected. These data were a subset of those collected from a prospective study conducted at the Diabetes Care Unit, Fondazione Policlinico Agostino Gemelli in Rome, Italy. The original protocol was approved by the local Ethics Committee (ref. 51266/19) and was conducted according to the Declaration of Helsinki. Each subject signed the appropriate informed consent form before any study activity was undertaken. Good metabolic control was defined as time in range ($70$-$180$ mg/dL) $>80\%$, which is better than currently recommended by guidelines. All subjects were being treated with metformin only. All CGM datasets were recorded between November 2021 and March 2022 with a Medtronic Guardian Connect System (Medtronic, Northridge, CA). Data were then extracted into a .csv file format for analysis.\\

\noindent
In order to prevent the patient's glucose levels from being affected by macroscopic factors such as food intake, physical activity or strong emotions, which could consequently complicate the analysis, the data set was narrowed to the night-time hours (8 p.m-8a.m). \\

\noindent
The procedure used to identify the best model to fit each subject's CGM data and to estimate the corresponding vector of model parameters is detailed below. Graphical representations and data analysis were computed using Matlab~\cite{matlab} and R~\cite{r}, respectively.

\bigskip

%%%%%%%%%%%%%%%%%%%%%%
\subsection{Model Description}
\label{Ordinary Differential Equations Models}

\noindent
As stated at the end of section \ref{Introduction}, we developed our new model
by generalizing the following ODE proposed by Sakulrang et al.~\cite{Sakulrang1}:
\begin{equation}
\frac{dG}{dt}=k_{GX}-k_{XG}G(t), \ G(0)=G_b,
\end{equation}

\noindent
where $G(t)$ represents blood glucose concentration at time $t$, $k_{GX}$ is a constant representing the entry rate of glucose into the blood, $k_{XG}$ is a constant representing the elimination rate of glucose from the blood and $G_b$ represents the basal glycemia.\\

\noindent
Our proposed new model consists of the system of three first-order differential equations shown in Eq.~(\ref{models}), where $G(t)$ is blood glucose concentration, $H(t)$ represents a rate of glucose elimination from the blood, and $Y(t)$ represents random changes in elimination rate due to random changes in the body of a patient. The parameter $k_G$ represents a constant rate of glucose inflow into the blood from the surrounding tissue and $k_H$ and $k_{XH}$ are constants representing rates of increase and decrease of $H$.  The symbol $\delta (\cdot)$ indicates the Dirac delta function. The detailed meanings of all variables and parameters are given in Tables~\ref{StateVariables} and~\ref{Parameters}.

\begin{equation}
\left\{
\begin{array}{rl}
\displaystyle \frac{dG}{dt}&= k_G-H(t)G(t) \\
\rule{0cm}{0.7cm}\displaystyle \frac{dH}{dt}&=k_H-k_{XH}H(t)+Y(t) \\
\rule{0cm}{0.7cm}\displaystyle \frac{dY}{dt}&=\sum_{i=0}^{N_Y}Y_i\delta (t-t_i)\\
\rule{0cm}{0.4cm}G(0)&=G_0\\
H(0)&=H_0\\
Y(0)&=Y_0
\end{array}
\right.
\label{models}
\end{equation}

\noindent
In the $G$ equation, we have made the assumption of linearity
between $G$ and $\frac{dG}{dt}$ as it is the simplest possible assumption to
represent the variation of a quantity in time.  We have made a similar linearity assumption
in the $H$ equation.
For the $Y(t)$ equation, we note that in subsection~\ref{Collection of Data}, we mentioned that the data set was restricted to the night-time hours (8 p.m-8 a.m) in order to reduce the impact of major blood glucose fluctuations induced by life events such as meals and sustained high-level physical activity. However, as shown in the data plots in section~\ref{Results and Discussion}, a considerable degree of variation in the levels of blood glucose was still recorded. This led to the hypothesis that there were intrinsic unknown factors determining the observed glucose fluctuations. Such perturbations did not translate into immediate discontinuities in the glycemic levels, but rather into relatively smooth oscillations. Therefore, the choice was made to model the random variations by introducing the perturbation term $Y(t)$ in the equation for the variation of $H$ as a piecewise constant function, with jumps at apparently random times $t_0,t_1,\ldots, t_{N_Y}$ and corresponding jump intensities $Y_0, Y_1,\ldots,Y_{N_Y}$, which can assume either positive or negative values. \\

\bigskip

\begin{table}[h!]
\begin{center}
\begin{tabular}{||p{3.5cm}|p{6cm}|p{3.5cm}||}
\hline
\textbf{State Variables} & \textbf{Definition} & \textbf{Unit of Measurement} \\
\hline
$G(t)$ & Blood glucose concentration. & $[mg/dL]$\\
\hline
$H(t)$ & Rate of glucose elimination from blood into the external environment. & $[min^{-1}]$ \\
\hline
$Y(t)$ & Perturbation of the rate of change of the  $H(t)$ function. & $[min^{-2}]$ \\
\hline
\end{tabular}
\end{center}
\caption{State Variables.}
\label{StateVariables}
\end{table}

\begin{center}
\tablefirsthead{%
\hline
\multicolumn{1}{||l}{\textbf{Parameters}} &
\multicolumn{1}{|l|}{\textbf{Definition}} &
\multicolumn{1}{|l|}{\textbf{Unit of Measurement}} &
\multicolumn{1}{|l|}{\textbf{Ranges}} \\
\hline}
\tablehead{%
\hline
\multicolumn{4}{||l||}{\small\sl continued from previous page}\\
\hline
\multicolumn{1}{||l}{\textbf{Parameters}} &
\multicolumn{1}{|l|}{\textbf{Definition}} &
\multicolumn{1}{|l|}{\textbf{Unit of Measurement}} &
\multicolumn{1}{|l|}{\textbf{Ranges}} \\
}
\tabletail{%
\hline
\multicolumn{4}{|l|}{\small\sl continued on next page}\\
\hline}
\tablelasttail{\hline}
\bottomcaption{Parameters.}
\begin{supertabular}{||p{2cm}|p{3.8cm}|p{2cm}|p{2cm}||}
$k_G$ & Constant entry rate of glucose into the bloodstream. & $[mg/dL\ min^{-1}]$ & $[-10^6,10^6]$\\
\hline
$k_H$ & Constant rate of change of $H$. & $[min^{-2}]$ & $[-10^6,10^6]$\\
\hline
$k_{XH}$ & Constant rate of change of $H$ per minute.  & $[min^{-1}]$ & $[0.0001,0.2]$\\[5mm]
\hline
$N_Y$ & Number of external/internal shocks caused by unknown external/internal factors. & $ -$ & $[1,40]$\\
\hline
$Y_i, \forall i \in S\subseteq \mathbb{N};\ |S|=N_Y <\infty.$ & Intensity of external/internal shocks due to unknown external/internal factors. & $[min^{-1}]$ & $ [-0.01,0.01]$\\
\hline
$t_i, \forall i \in S\subseteq \mathbb{N};\ |S|=N_Y <\infty. $ & Times at which shocks occur. & $[min]$ & $[0,715]$\\
\hline
$\Delta t_i = t_{i+1}-t_i$&Time between shocks&$[min]$& \\
\hline
$G_0$ & Initial blood glucose concentration (more specifically, glycemic value in resting conditions, also known as basal or resting glycemia). & $[mg/dL]$ & $[30,180]$\\
\hline
$H_0$ & Initial rate of glucose elimination from blood into the external environment. & $[min^{-1}]$ & $[0.0001,0.2]$\\
\hline
$Y_0$ & Initial rate of change of the first derivative of the $H(t)$ function. & $[min^{-2}]$ & $[-1,1]$ \\
\hline
\end{supertabular}
\label{Parameters}
\end{center}

\noindent
We have created a family of eight mathematical models of the form of Eq.~(\ref{models})
for the eight patients, indexed by the number $N_Y$ of jumps, where an increase by one jump corresponds to two additional parameters in the model (the timing and the intensity of the new jump).\\

%We may assume that these waiting times between jumps follow an exponential distribution of parameter $\lambda({N_Y})$: this assumption is actually not necessary for the adaptation of the model to data from a given patient, but will be useful after the fit to summarize the results.

\bigskip
\noindent
It is clear that for any given disturbance function $Y(t) = Y_0 + \sum_{i : \, t_i \leqslant t} Y_i$, the rate of glucose elimination from blood, $H(t)$, tends asymptotically to a value determined by $Y(t)$, which in general is different from the steady-state solution $\frac{k_H}{k_{XH}}$.\\

\bigskip
\noindent
If we assume that at the time $t_0=0$, $\frac{dG}{dt}=0$ and $\frac{dH}{dt}=0$, then $k_G=H_0G_0$ ($G_0$ known). If $H_0$ is estimated, then it follows that $k_H=k_{XH}H_0-Y_0$ is known. Therefore, $k_G$ and $k_H$ are determined parameters whereas the remaining ones are free. Thus the vector of parameters which needs to be estimated is the following:

\begin{equation}
\theta=[k_{XH},H_0,Y_0, (Y_i)_{i=1,\ldots, N_Y}, (t_i)_{=1,\ldots, N_Y}].
\end{equation}

%%%%%%%%%%%%%%%%%%%%%%%%%%%%%%%%%%%%%%%%%%
\subsection{Exact Solution}
\label{Exact Solution}

\noindent
In Eq.~(\ref{models}), we first noted that the solution of the $Y(t)$ equation is a piecewise-constant function with the values $Y(t) = \sum_{i=0}^n\,Y_i$,
for $t_n \leq t < t_{n+1}$. In order to solve the system analytically Eq.~(\ref{models}), a number of jumps equal to 2 was initially assumed ($N_Y=2$):

\begin{equation}
\left\{
\begin{array}{rl}
\displaystyle \frac{dG}{dt}&= k_G-H(t)G(t) \\
\rule{0cm}{0.7cm}\displaystyle \frac{dH}{dt}&=k_H-k_{XH}H(t)+Y(t) \\
\rule{0cm}{0.7cm}\displaystyle \frac{dY}{dt}&=\sum_{i=0}^{2}Y_i\delta (t-t_i)\\
\rule{0cm}{0.4cm}G(0)&=G_0\\
H(0)&=H_0\\
Y(0)&=Y_0
\end{array}
\right.
\label{models1}
\end{equation}

\noindent
Then, the procedure was generalized to any value of $N_Y$.\\

\noindent
Firstly, we considered the second equation of the system (\ref{models1}) for $0\leq t<t_1$:\\

$$\frac{dH}{dt}=k_H-k_{XH}H(t)+Y_0$$

\noindent
whose solution is

$$H(t)=c_1^1e^{-k_{XH}t}+c_2^1$$

\noindent
with $c_2^1=\frac{Y_0+k_H}{k_{XH}}$ and $c_1^1=H_0-c_2^1$. \\
\noindent
Secondly, we observed that for for the second time interval $t_1\leq t<t_2$, we had $Y(t)=Y_0+Y_1$ and therefore:

$$\frac{dH}{dt}=k_H-k_{XH}H(t)+Y_0+Y_1.$$

\noindent
Consequently, the functional form of $H$ was similar to the previous one:

$$H(t)=c_1^2e^{-k_{XH}t}+c_2^2$$

\noindent
where, for this interval, $c_2^2=\frac{Y_0+Y_1+k_H}{k_{XH}}$ and $c_1^2$ generic constant. In order to find $c_1^2$, we imposed the continuity condition for $H$ at $t_1$ that $H(t_1^+)=H(t_1^-)$ from which we deduced that:

$$c_1^1e^{-k_{XH}t_1}+c_2^1=c_1^2e^{-k_{XH}t_1}+c_2^2 \ \Rightarrow \ c_1^2=\frac{c_1^1e^{-k_{XH}t_1}+c_2^1-c_2^2}{e^{-k_{XH}t_1}}.$$

\noindent
In this way, we found constants that guarantee the continuity of $H(t)$.
Naturally, we applied the same reasoning for $t_2\leq t<t_{final}$ and we got:

$$H(t)=c_1^3e^{-k_{XH}t}+c_2^3$$

\noindent
with $c_2^3=\frac{Y_0+Y_1+Y_2+k_H}{k_{XH}}$ and $c_1^3=\frac{c_1^2e^{-k_{XH}t_2}+c_2^2-c_2^3}{e^{-k_{XH}t_2}}$. 		\\

\noindent
Indicating with $n=1,\ldots,\ N_Y+1$ the number of intervals generated by the time jumps and with $H_n$ the functional form of $H$ in the $n-$th interval, we deduced that

$$H_n(t)=c_1^n e^{-k_{XH}t} +c_2^n$$

\noindent
with $c_2^n=\frac{\sum_{i=0}^{n-1}Y_i+k_H}{k_{XH}}$ and $c_1^n=\frac{c_1^{n-1}e^{-k_{XH}t_{n-1}}+c_2^{n-1}-c_2^n}{e^{-k_{XH}t_{n-1}}}$ if $t\in [t_{n-1},t_{n}[$ and $n=1,\ldots,\ N_Y+1$. In this context we imposed $t_0=0$ and $t_{N_Y+2}=t_{final}$.\\
\vspace{0.5 cm}

\noindent
Using the above solutions for the $H$ equation in Eq.~(\ref{models}), we could now solve the
glycemia equations $G(t)$ in each interval $[t_{n-1},t_{n}]$ which were of the form:

\begin{equation}\label{Ginter1}
\left\{
\begin{array}{rl}
\displaystyle \frac{dG}{dt}&= k_G-(c_1^n e^{-k_{XH}t}+c_2^n)G(t)\\
\rule{0cm}{0.4cm}G(t_{n-1}^+)&=G(t_{n-1}^-)\\
\end{array}
\right.
\end{equation}

\noindent
with $c_i^n, \ i=1,2$ and the initial condition that depended on the interval in which $t$ was located. Naturally, $t_0=0,\ G(t_{0}^-)=G_0$. Consequently, $G(0)=G_0$ was the initial condition for $n=1$. Eq.~(\ref{Ginter1}) is a first order linear differential equation with non-constant coefficients. For ease of writing, we rewrote Eq.~(\ref{Ginter1}) in the simpler form

\begin{equation}\label{Ginter2}
\left\{
\begin{array}{rl}
\displaystyle \frac{dy}{dx}&= a(x)y(x)+b(x)\\
\rule{0cm}{0.4cm} y(x_0)&=y_0\\
\end{array}
\right.
\end{equation}

\noindent
We then followed the well-known method of solution to solve this equation and got:

\begin{equation}
\begin{array}{rl}
\displaystyle y(x)&=e^{\int_{x_0}^x a(s)ds}[ y_0+\int_{x_0}^x b(t)e^{-\int_{x_0}^t a(s)ds}dt]=\\
\rule{0cm}{0.8cm}&=y_0e^{\int_{x_0}^x a(s)ds}+\int_{x_0}^x b(t)e^{\int_{x_0}^x a(s)ds}e^{-\int_{x_0}^t a(s)ds}dt=\\
\rule{0cm}{0.8cm}&=y_0e^{\int_{x_0}^x a(s)ds}+\int_{x_0}^x b(t)e^{\int_{x_0}^t a(s)ds}e^{\int_{t}^x a(s)ds}e^{\int_{x_0}^t -a(s)ds}=\\
\rule{0cm}{0.8cm}&=y_0e^{\int_{x_0}^x a(s)ds}+\int_{x_0}^x b(t)e^{\int_{t}^x a(s)ds}dt
\end{array}
\end{equation}

\noindent
In the original notation, the solution for $G(t)$ in each interval was as follows

\begin{equation}
\begin{array}{rl}
\displaystyle G(t)=G(t_{n-1}^-)e^{\int_{t_{n-1}^+}^t -(c_1^n e^{-k_{EH}s}+ c_2^n)ds}+&\int_{t_{n-1}^+}^t k_G e^{\int_{p}^t -(c_1^n e^{-k_{EH}s}+ c_2^n)ds}dp \ ,\\
\rule{0cm}{0.8cm}& t \in [t_{n-1},t_n[
\end{array}
\end{equation}
%%%%%%%%%%%%%%%%%%%%%%%%%%%%%%%%%%%%%%%%%
\subsection{Numerical Methods}
\label{Numerical Methods}

\noindent
The data set utilized in this work consists of a total of 144 measurements of the interstitial glucose levels (approximately one observation every five minutes for 12 hours) for each of the eight patients. As already mentioned in subsection \ref{Collection of Data}, these data were collected from CGM devices during the time slot 8:00 PM - 8:00 AM. Depending on the value of $N_Y$, a family of models was obtained. The present work aimed at finding, for each experimental subject, which value of $N_Y$ gave rise to the model that best fits their CGM tracing according to Akaike's Information Criterion~\cite{AIC}.\\

\bigskip
\noindent
The procedure which has been followed for each experimental subject consists of several steps:

\begin{enumerate}[label=\Roman*.]
\item For each $N_Y$ the system of differential equations (\ref{models}) was solved analytically.

\item Parameter estimation was carried out by  Ordinary Least Squares (OLS). Specifically, the loss function

$$L=\sum_{i=1}^{144} (G_i(t_i)-\hat{G}_i(\theta,t_i))^2$$

\noindent
was minimized, where $G_i(t_i)$ represents the CGM value recorded at time $t_i$ and $\hat{G}_i(\theta,t_i)$ is the corresponding estimated value obtained by our model. For the minimization, the Particle Swarm Optimization (PSO) ~\cite{pso2,pso3,pso4} global method was used. It should be noted that in preliminary runs the Nelder-Mead local optimization algorithm was used. However, due to its poor performance (inability to vary the time-points of the jumps due to essentially constant values of the loss function around each local optimum) we opted to use the Particle Swarm Optimization method~\cite{pso2,pso3,pso4}.

\item Akaike Information Criterion (AIC) values were calculated for all $N_Y$ values tested and the model with the smallest AIC value was picked as the "best" model for that subject. For comparison, we also selected the "best" model using the Bayesian Information Criterion (BIC) value~\cite{BIC,AICVSBIC}.
\end{enumerate}

%%%%%%%%%%%%%%%%%%%%%%%%%%%%%%%%%%%%%%%%%%%%%%%%%%%%%%%%

\section{Results and Discussion}
\label{Results and Discussion}

\begin{figure}[H]
  \begin{minipage}[b]{0.4\linewidth}
    \includegraphics[width=8cm, height=6cm]{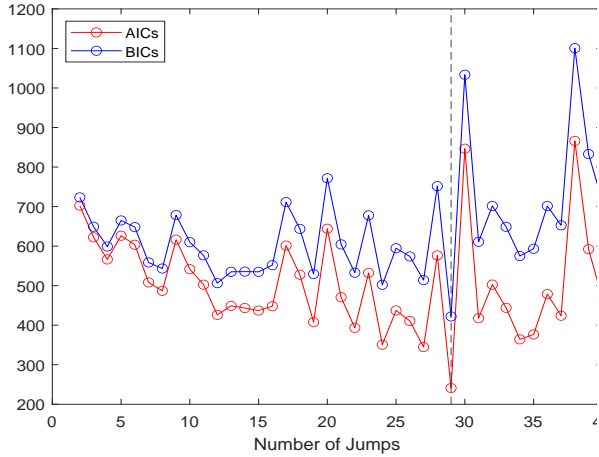}
\subcaption{AIC and BIC trends over the number of jumps.}
\vspace{4ex}
\label{AA01a}
   \end{minipage}
\hfill
   \begin{minipage}[b]{0.4\linewidth}
    \includegraphics[width=8cm,height=6cm]{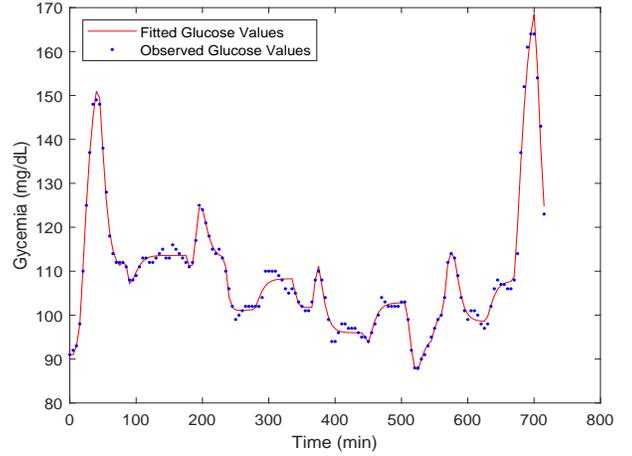}
\subcaption{Gycemia over time, $N_Y=29$.}
\vspace{5.7ex}
\label{AA01b}
   \end{minipage}
\hfill
   \begin{minipage}[b]{0.4\linewidth}
    \includegraphics[width=8cm,height=6cm]{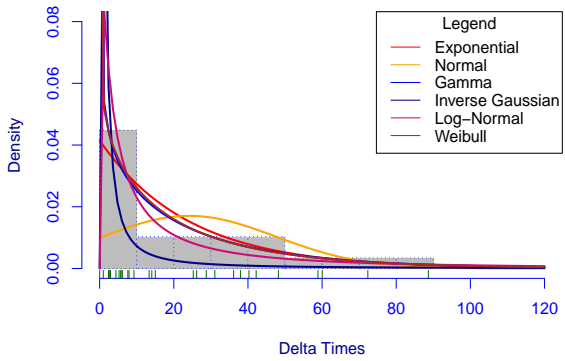}
\subcaption{Histogram of the Delta Times $(\Delta t_i)_{i=1\ldots,N_Y-1}$.}
\vspace{4ex}
\label{AA01c}
   \end{minipage}
  \hfill
   \begin{minipage}[b]{0.4\linewidth}
    \includegraphics[width=8cm,height=6cm]{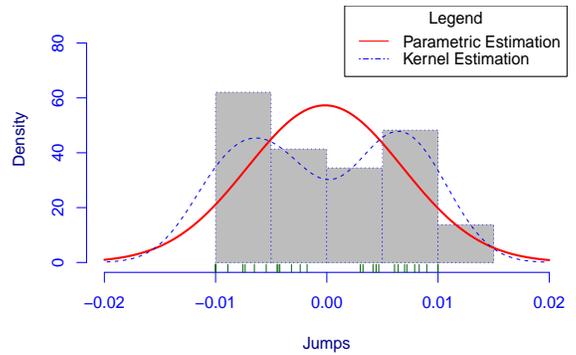}
\subcaption{Estimation of the density of the jumps values.}
\vspace{4ex}
\label{AA01d}
   \end{minipage}
\caption{Results from Patient AA01.}
\label{ResultsSubjectAA01}
\end{figure}

\begin{figure}[H]
  \begin{minipage}[b]{0.4\linewidth}
    \includegraphics[width=8cm, height=6cm]{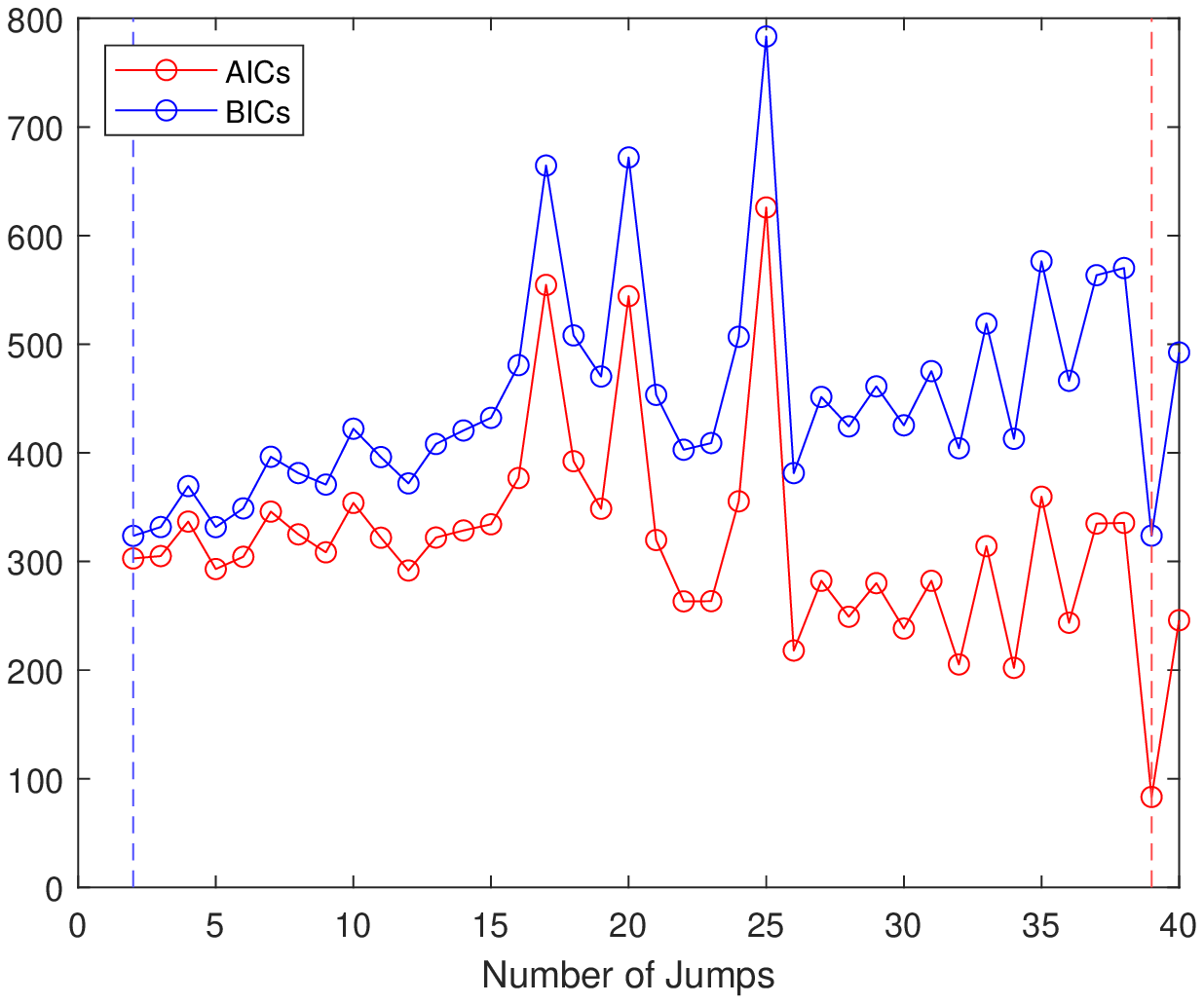}
\subcaption{AIC and BIC trends over the number of jumps.}
\label{BO01a}
\vspace{4ex}
   \end{minipage}
\hfill
   \begin{minipage}[b]{0.4\linewidth}
    \includegraphics[width=8cm,height=6cm]{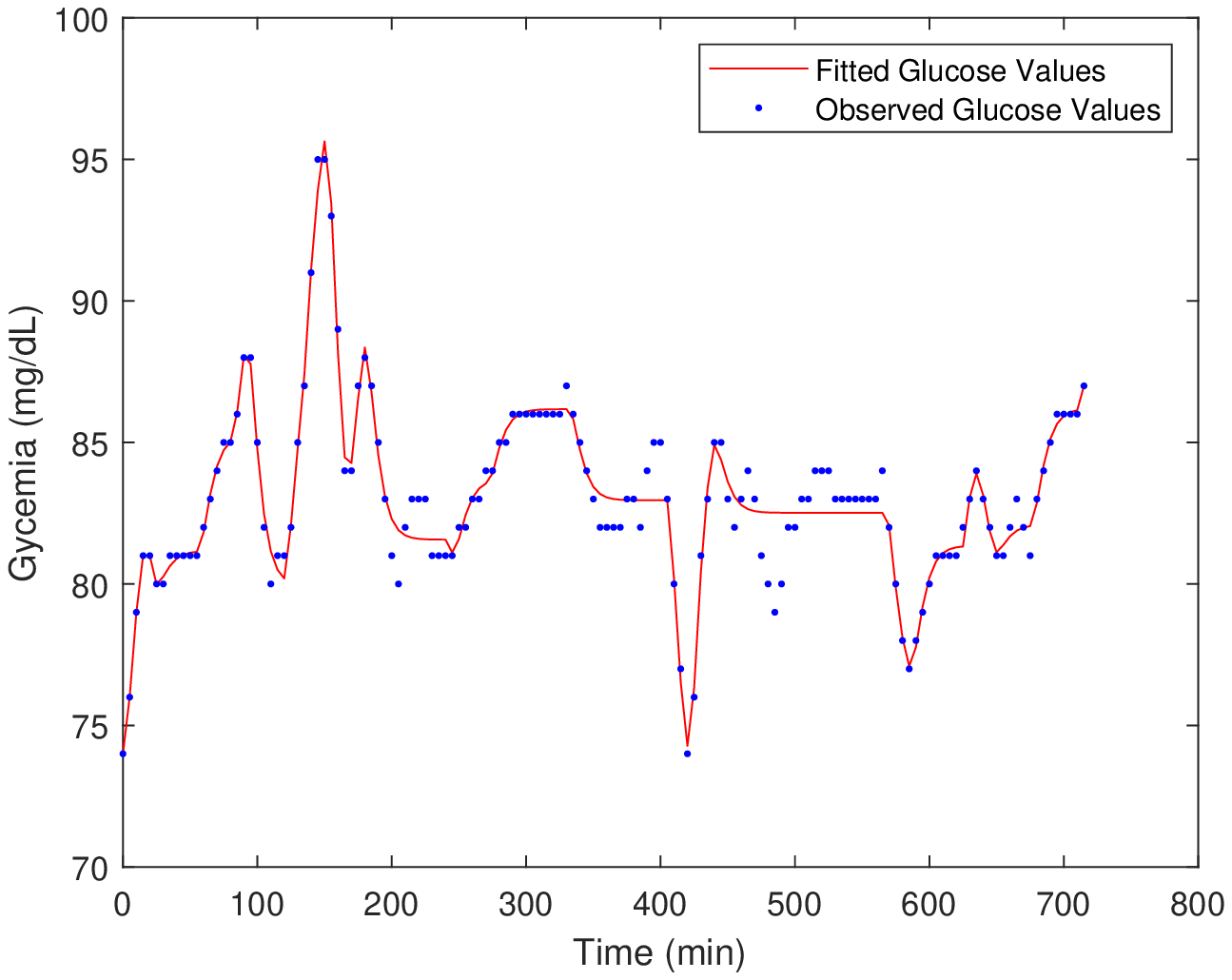}
\subcaption{Gycemia over time, $N_Y=39$.}
\vspace{5.7ex}
\label{BO01b}
   \end{minipage}
\hfill
   \begin{minipage}[b]{0.4\linewidth}
    \includegraphics[width=8cm,height=7cm]{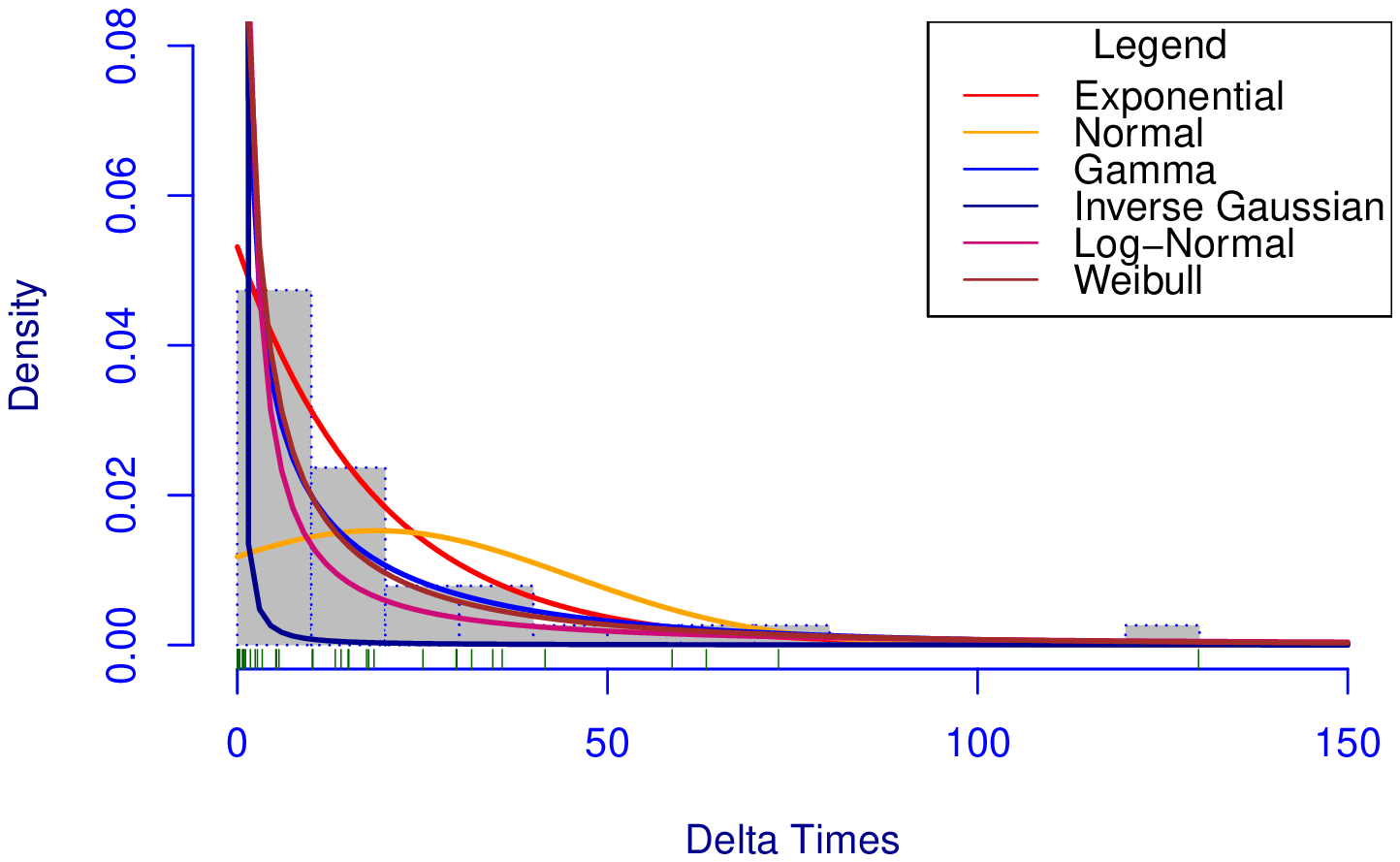}
\subcaption{Histogram of the Delta Times $(\Delta t_i)_{i=1\ldots,N_Y-1}$.}
\vspace{4ex}
\label{BO01c}
   \end{minipage}
  \hfill
   \begin{minipage}[b]{0.4\linewidth}
    \includegraphics[width=8cm,height=7cm]{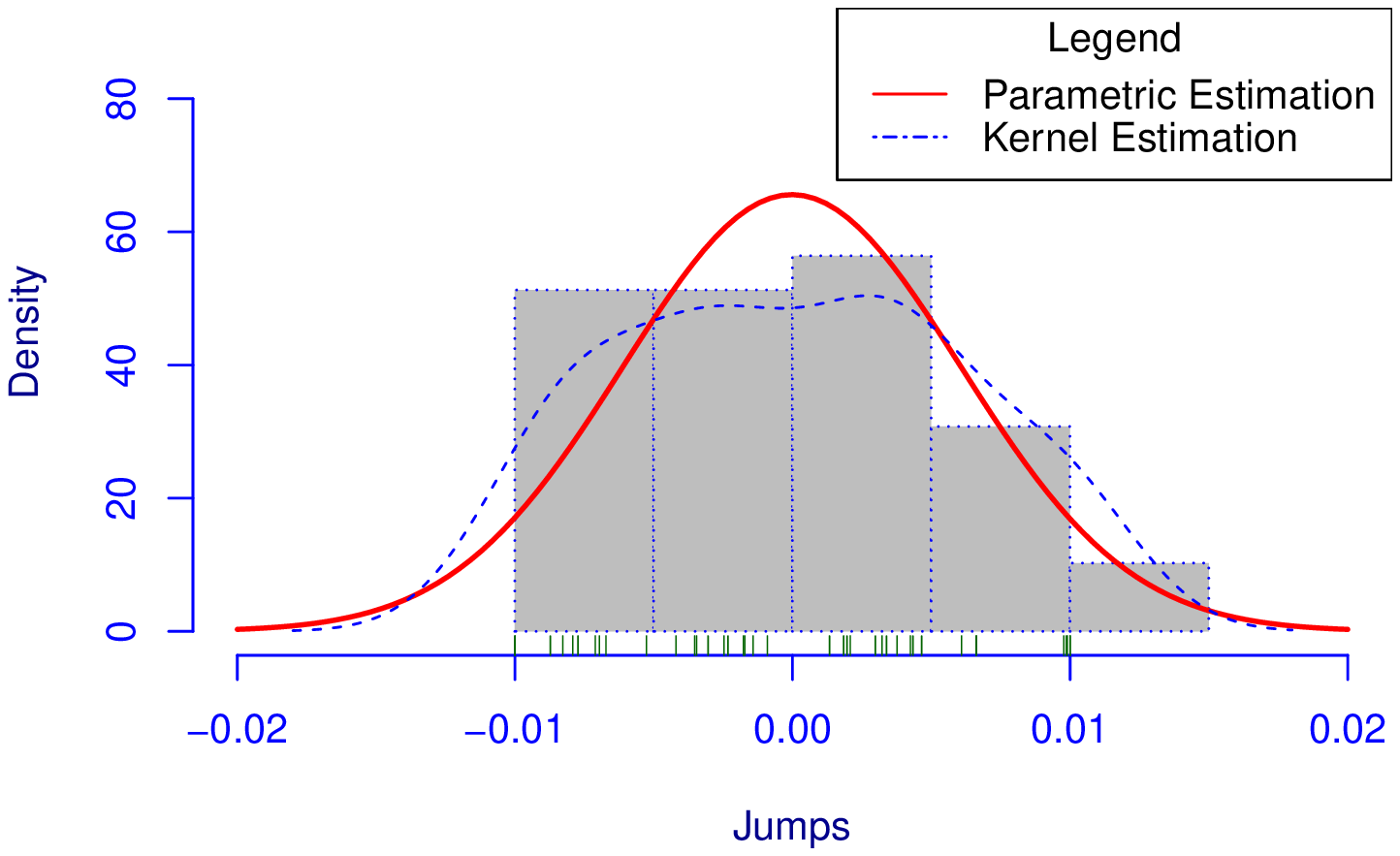}
\subcaption{Estimation of the density of the jumps values.}
\vspace{4ex}
\label{BO01d}
   \end{minipage}
\caption{Results from Patient BO01.}
\label{ResultsSubjectBO01}
\end{figure}

\begin{figure}[H]
  \begin{minipage}[b]{0.4\linewidth}
    \includegraphics[width=8cm, height=6cm]{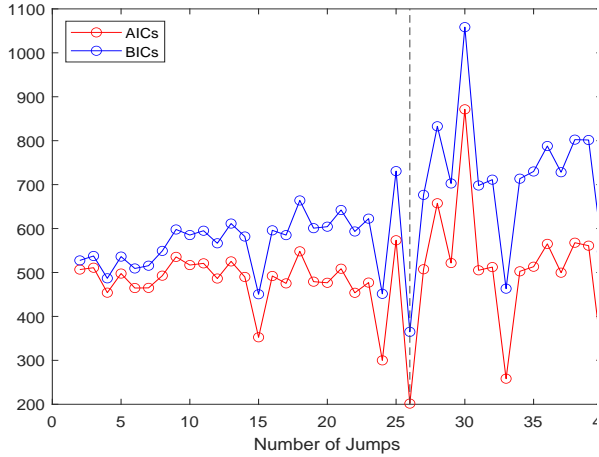}
\subcaption{AIC and BIC trends over the number of jumps.}
\vspace{4ex}
\label{CI01a}
   \end{minipage}
\hfill
   \begin{minipage}[b]{0.4\linewidth}
    \includegraphics[width=8cm,height=6cm]{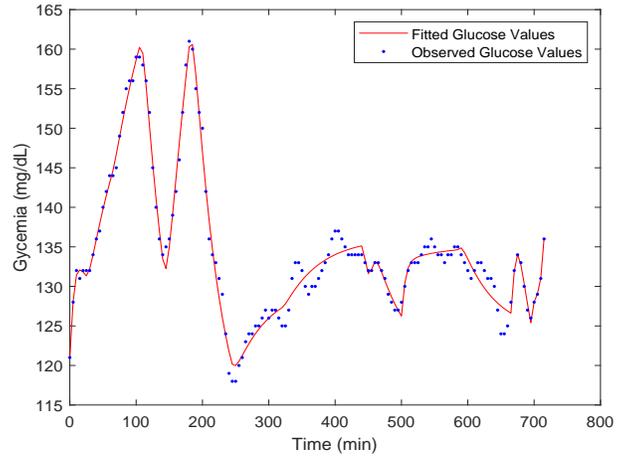}
\subcaption{Gycemia over time, $N_Y=26$}
\vspace{5.7ex}
\label{CI01b}
   \end{minipage}
\hfill
   \begin{minipage}[b]{0.4\linewidth}
    \includegraphics[width=8cm,height=7cm]{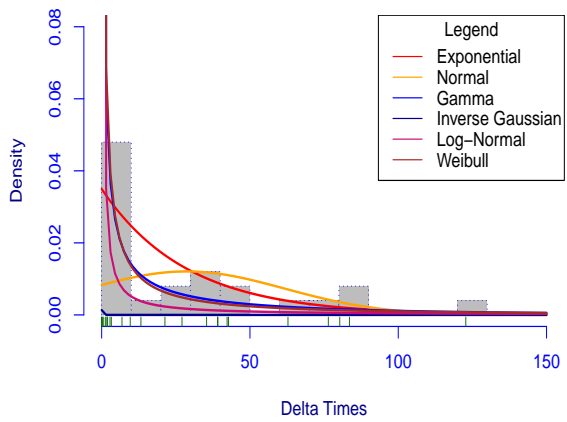}
\subcaption{Histogram of the Delta Times $(\Delta t_i)_{i=1\ldots,N_Y-1}$.}
\vspace{4ex}
\label{CI01c}
   \end{minipage}
  \hfill
   \begin{minipage}[b]{0.4\linewidth}
    \includegraphics[width=8cm,height=7cm]{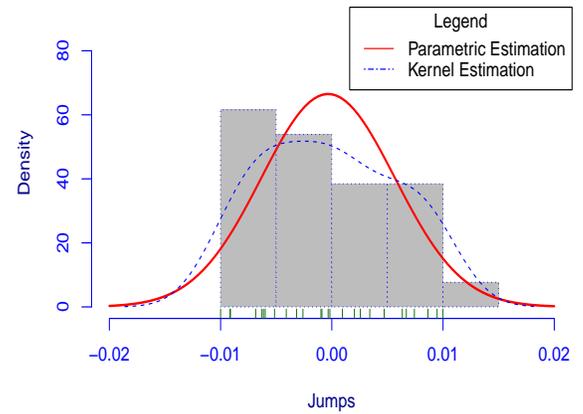}
\subcaption{Estimation of the density of the jumps values.}
\vspace{4ex}
\label{CI01d}
   \end{minipage}
\caption{Results from Patient CI01.}
\label{ResultsSubjectCI01}
\end{figure}

\begin{figure}[H]
  \begin{minipage}[b]{0.4\linewidth}
    \includegraphics[width=8cm, height=6cm]{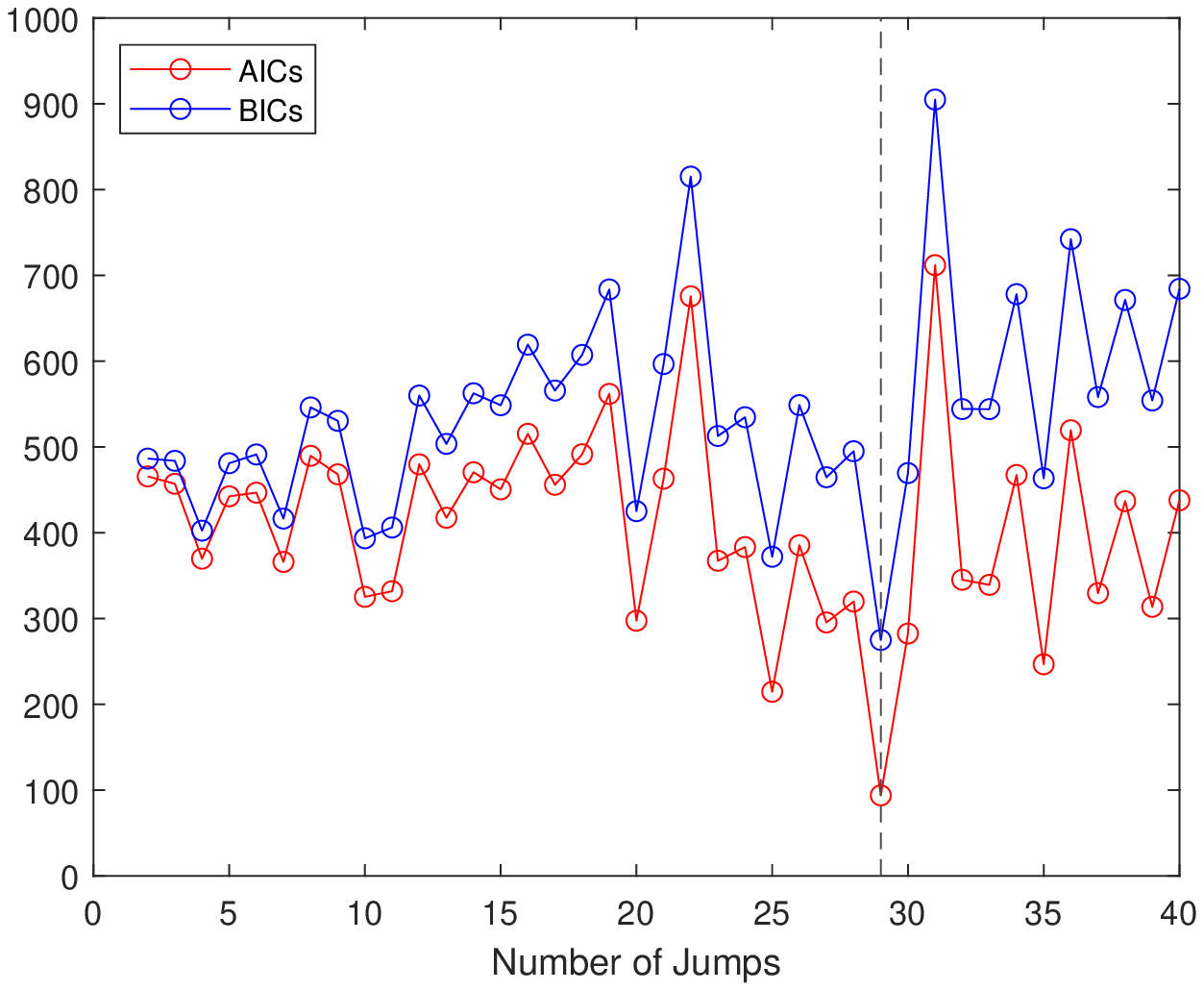}
\subcaption{AIC and BIC trends over the number of jumps.}
\vspace{4ex}
\label{LM01a}
   \end{minipage}
\hfill
   \begin{minipage}[b]{0.4\linewidth}
    \includegraphics[width=8cm,height=6cm]{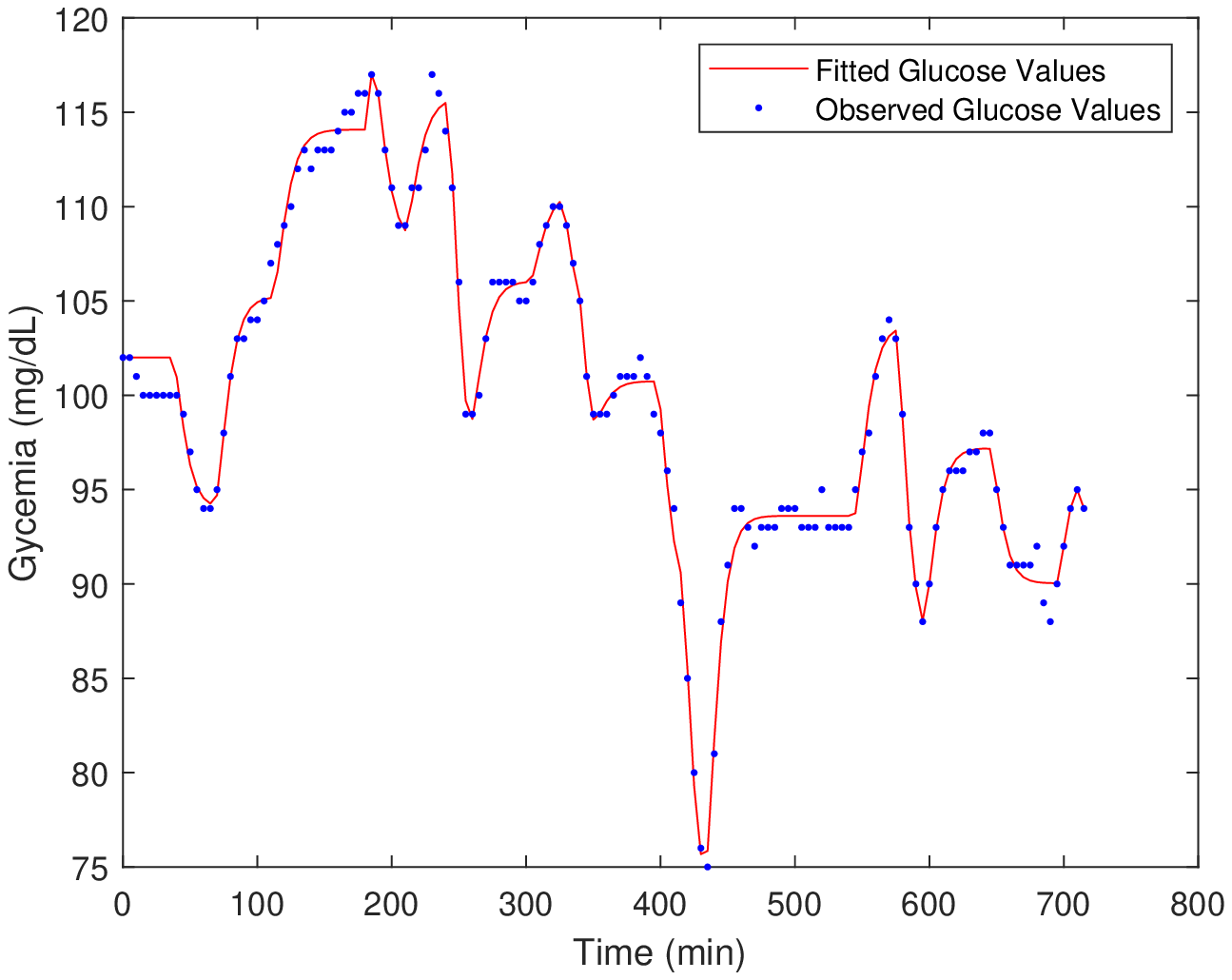}
\subcaption{Gycemia over time, $N_Y=29$.}
\vspace{5.7ex}
\label{LM01b}
   \end{minipage}
\hfill
   \begin{minipage}[b]{0.4\linewidth}
    \includegraphics[width=7.5cm,height=8cm]{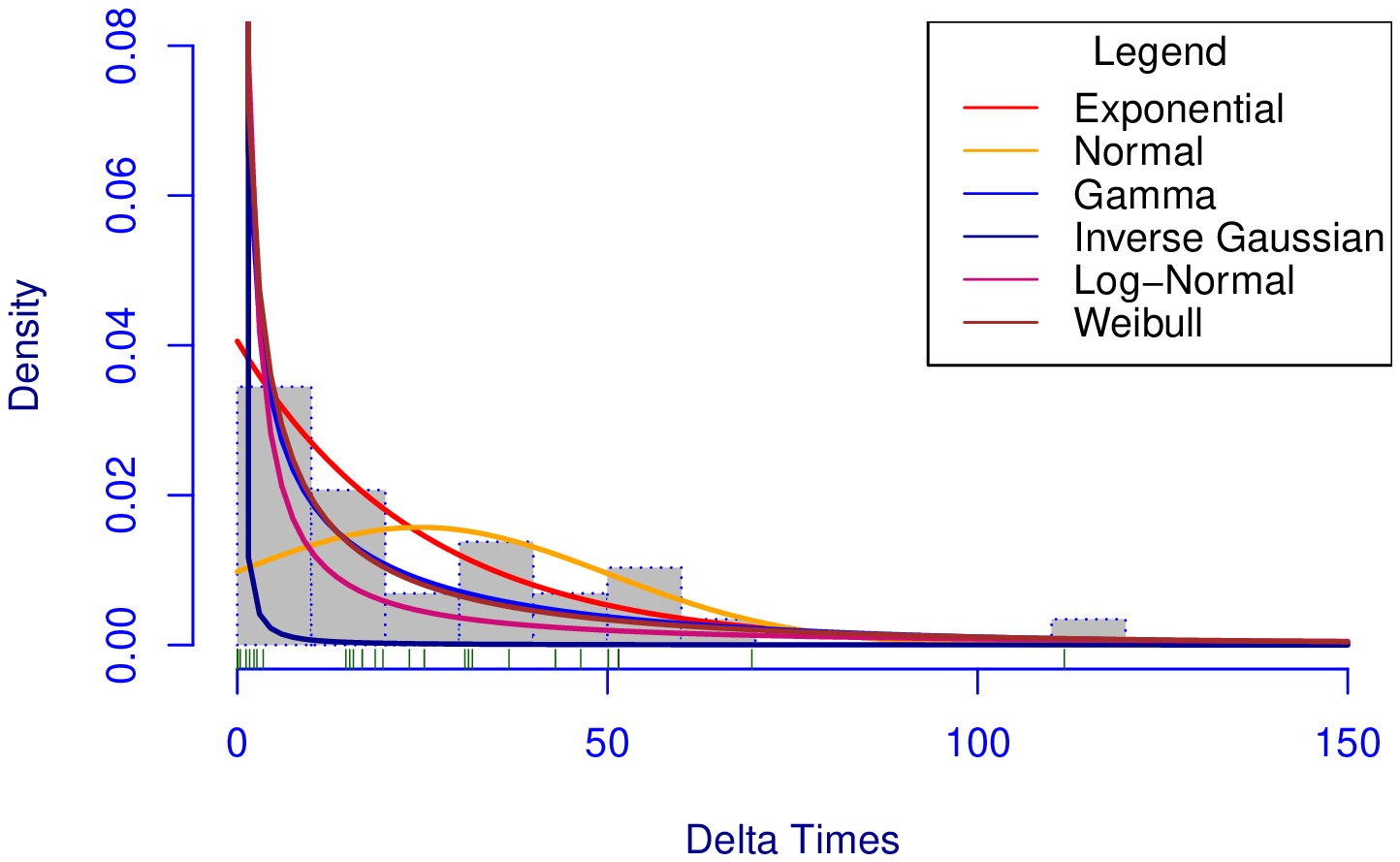}
\subcaption{Histogram of the Delta Times $(\Delta t_i)_{i=1\ldots,N_Y-1}$.}
\vspace{4ex}
\label{LM01c}
   \end{minipage}
  \hfill
   \begin{minipage}[b]{0.4\linewidth}
    \includegraphics[width=7.5cm,height=8cm]{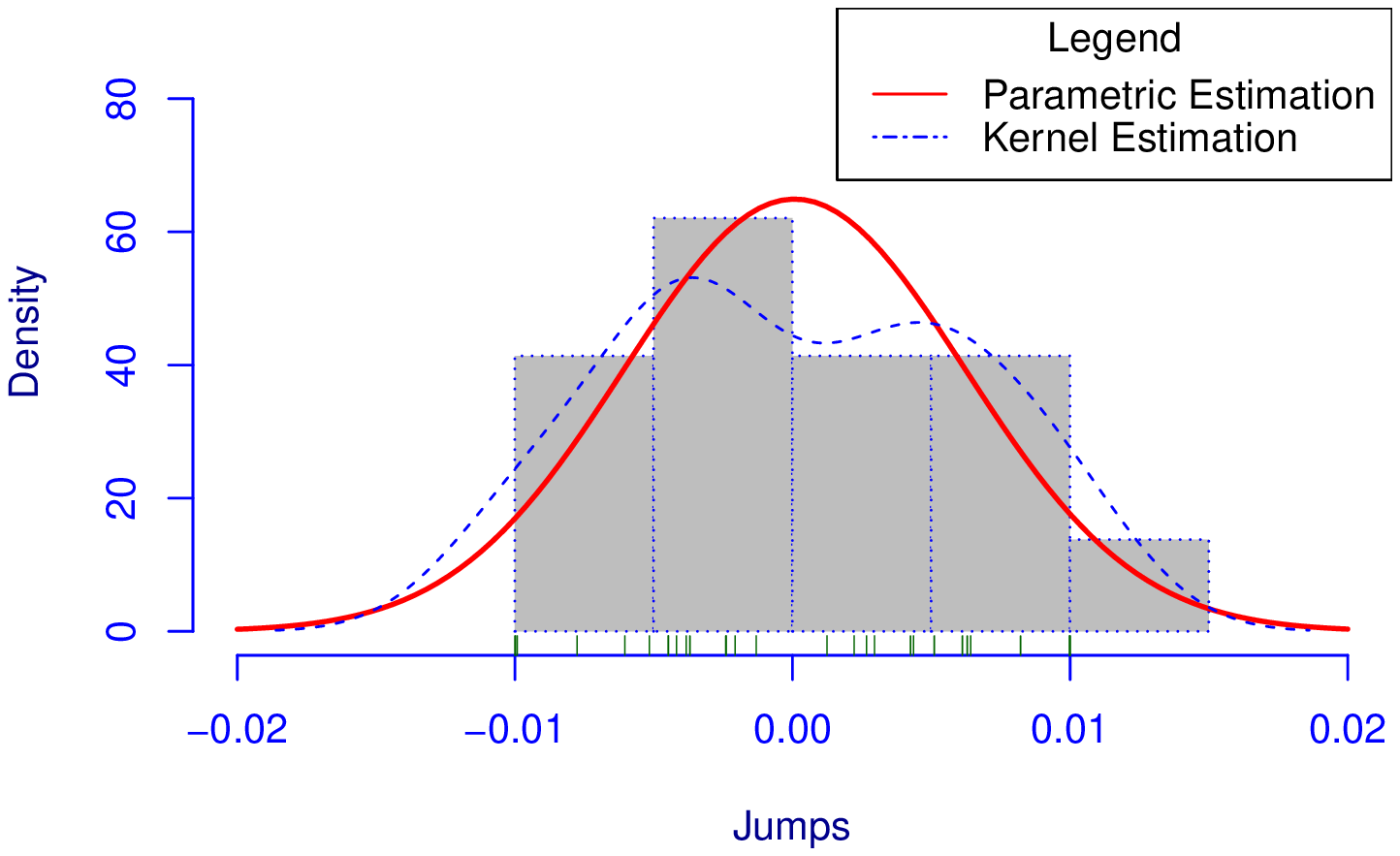}
\subcaption{Estimation of the density of the jumps values.}
\vspace{4ex}
\label{LM01d}
   \end{minipage}
\caption{Results from Patient LM01.}
\label{ResultsSubjectLM01}
\end{figure}

\begin{figure}[H]
  \begin{minipage}[b]{0.4\linewidth}
    \includegraphics[width=8cm, height=6cm]{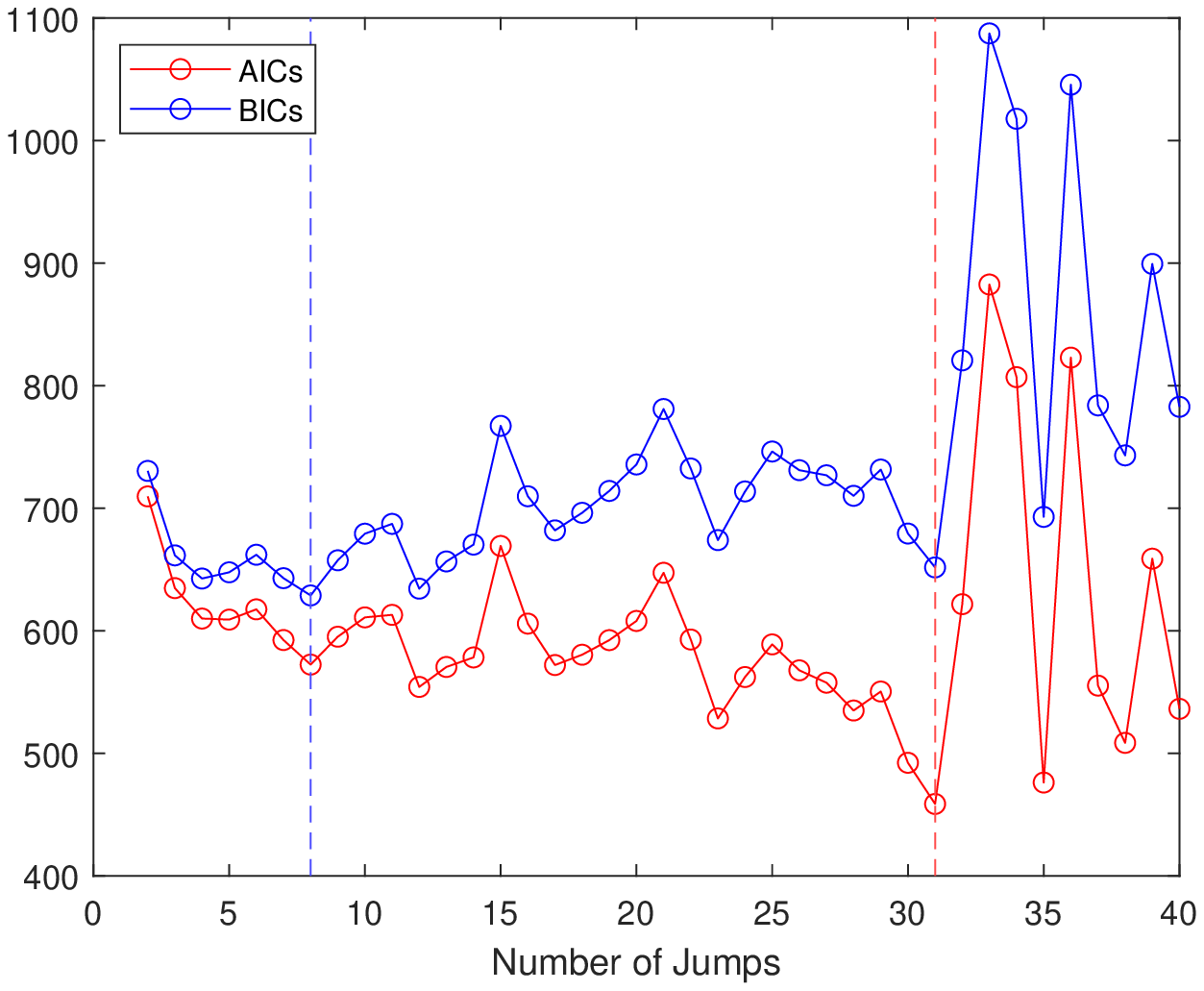}
\subcaption{AIC and BIC trends over the number of jumps.}
\vspace{4ex}
\label{LR01a}
   \end{minipage}
\hfill
   \begin{minipage}[b]{0.4\linewidth}
    \includegraphics[width=8cm,height=6cm]{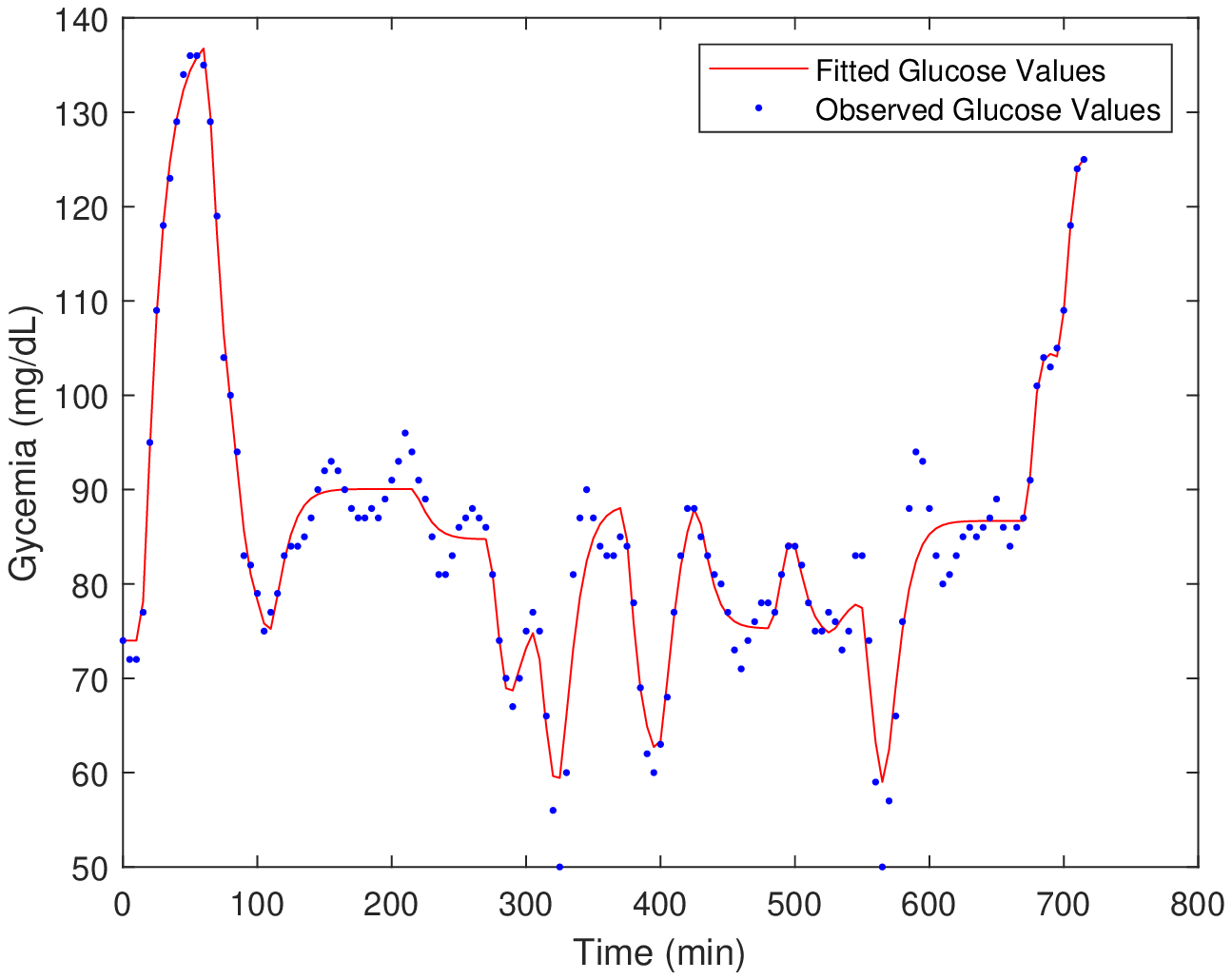}
\subcaption{Gycemia over time, $N_Y=31$.}
\vspace{5.7ex}
\label{LR01b}
   \end{minipage}
\hfill
   \begin{minipage}[b]{0.4\linewidth}
    \includegraphics[width=7.5cm,height=8cm]{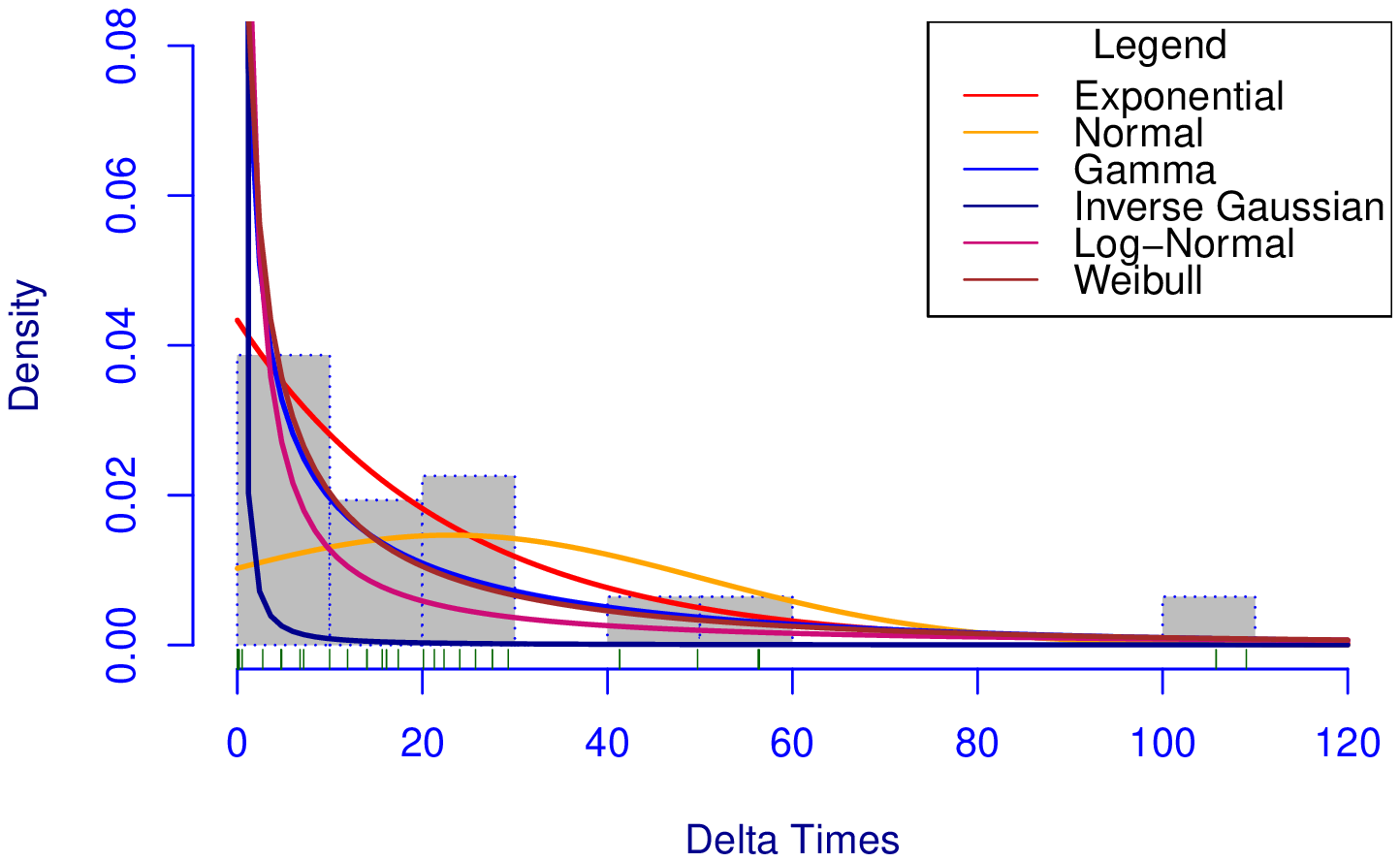}
\subcaption{Histogram of the Delta Times $(\Delta t_i)_{i=1\ldots,N_Y-1}$.}
\vspace{4ex}
\label{LR01c}
   \end{minipage}
  \hfill
   \begin{minipage}[b]{0.4\linewidth}
    \includegraphics[width=7.5cm,height=8cm]{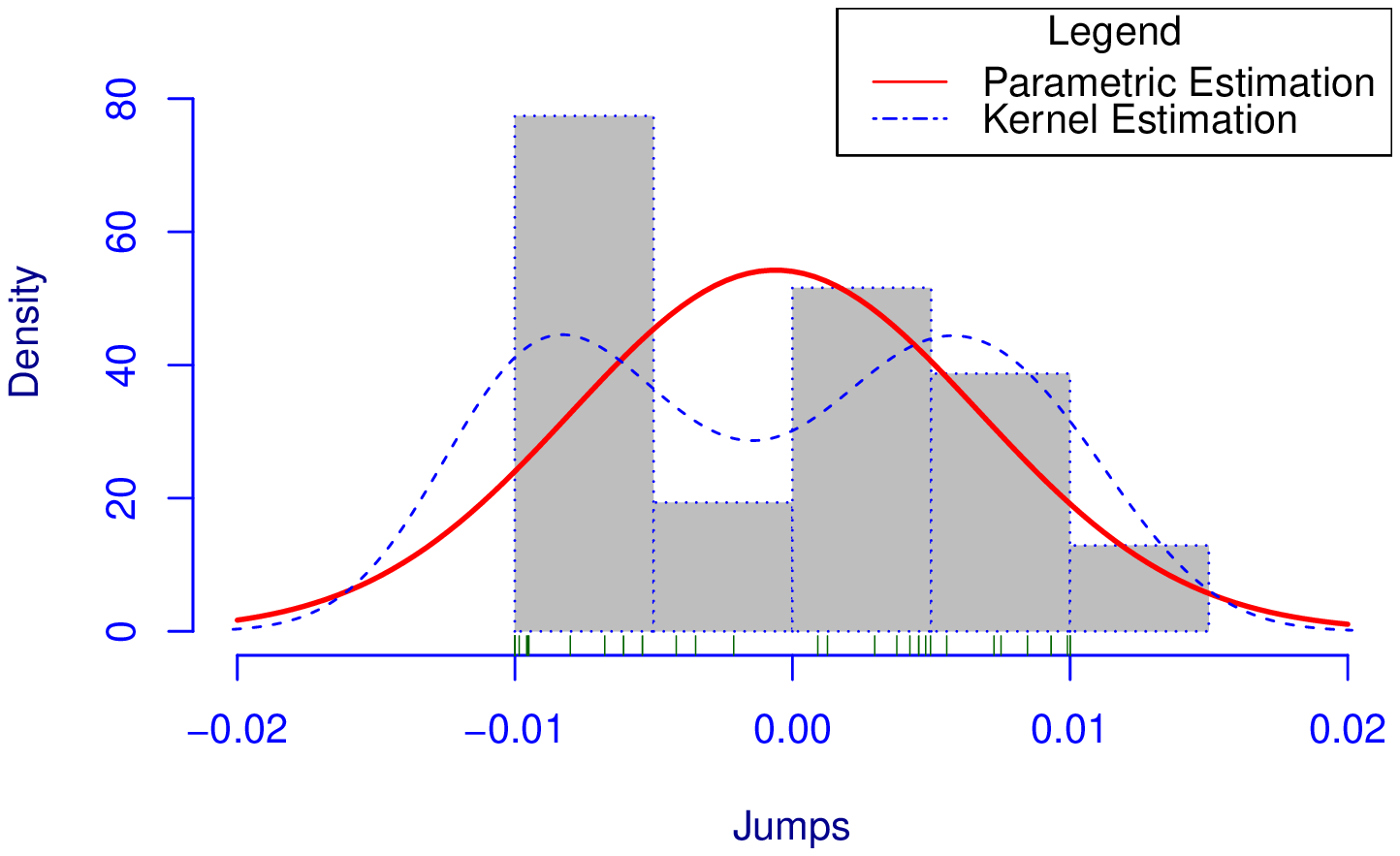}
\subcaption{Estimation of the density of the jumps values.}
\vspace{4ex}
\label{LR01d}
   \end{minipage}
\caption{Results from Patient LR01.}
\label{ResultsSubjectLR01}
\end{figure}

\begin{figure}[H]
  \begin{minipage}[b]{0.4\linewidth}
    \includegraphics[width=8cm, height=6cm]{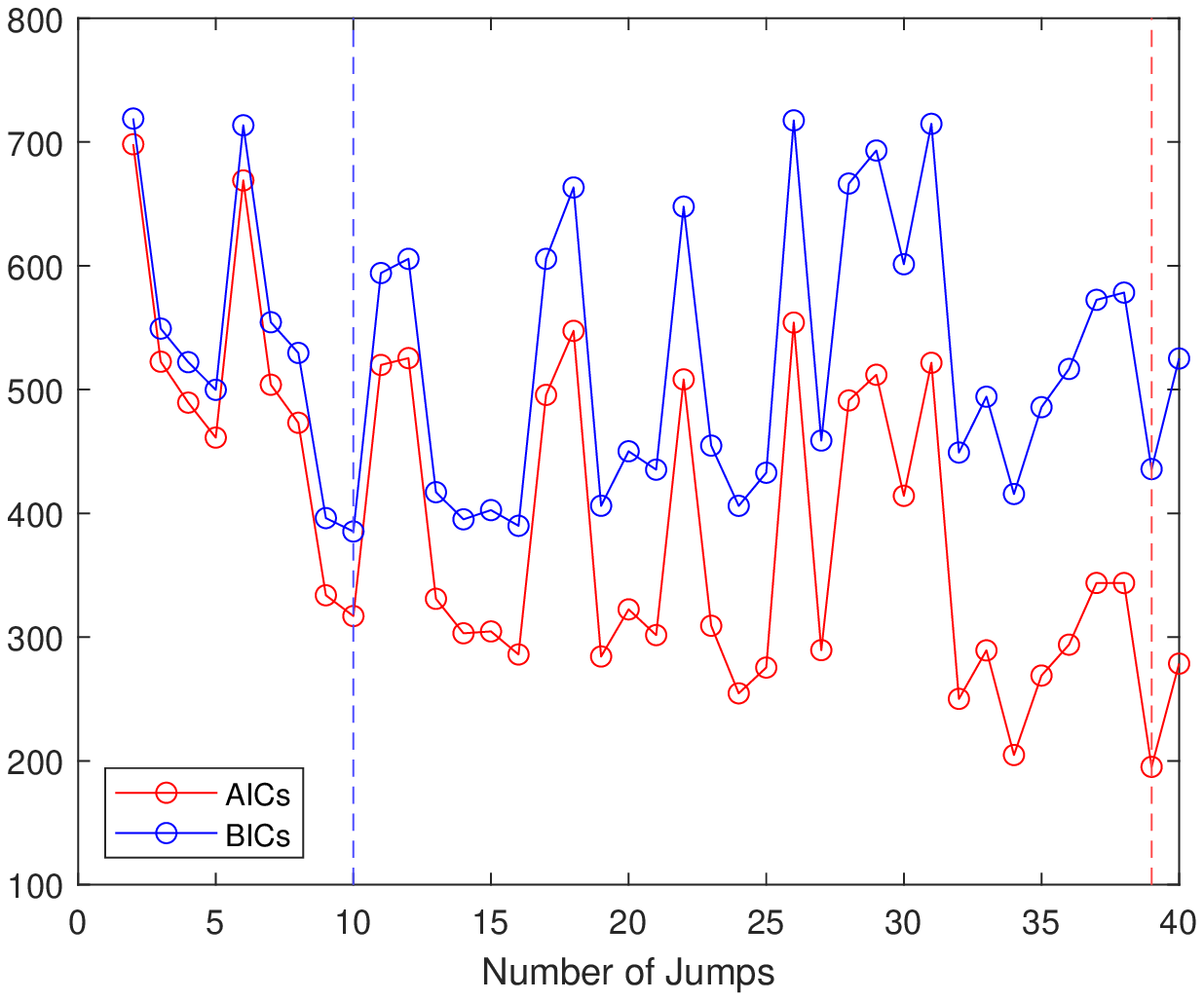}
\subcaption{AIC and BIC trends over the number of jumps.}
\vspace{4ex}
\label{MA01a}
   \end{minipage}
\hfill
   \begin{minipage}[b]{0.4\linewidth}
    \includegraphics[width=8cm,height=6cm]{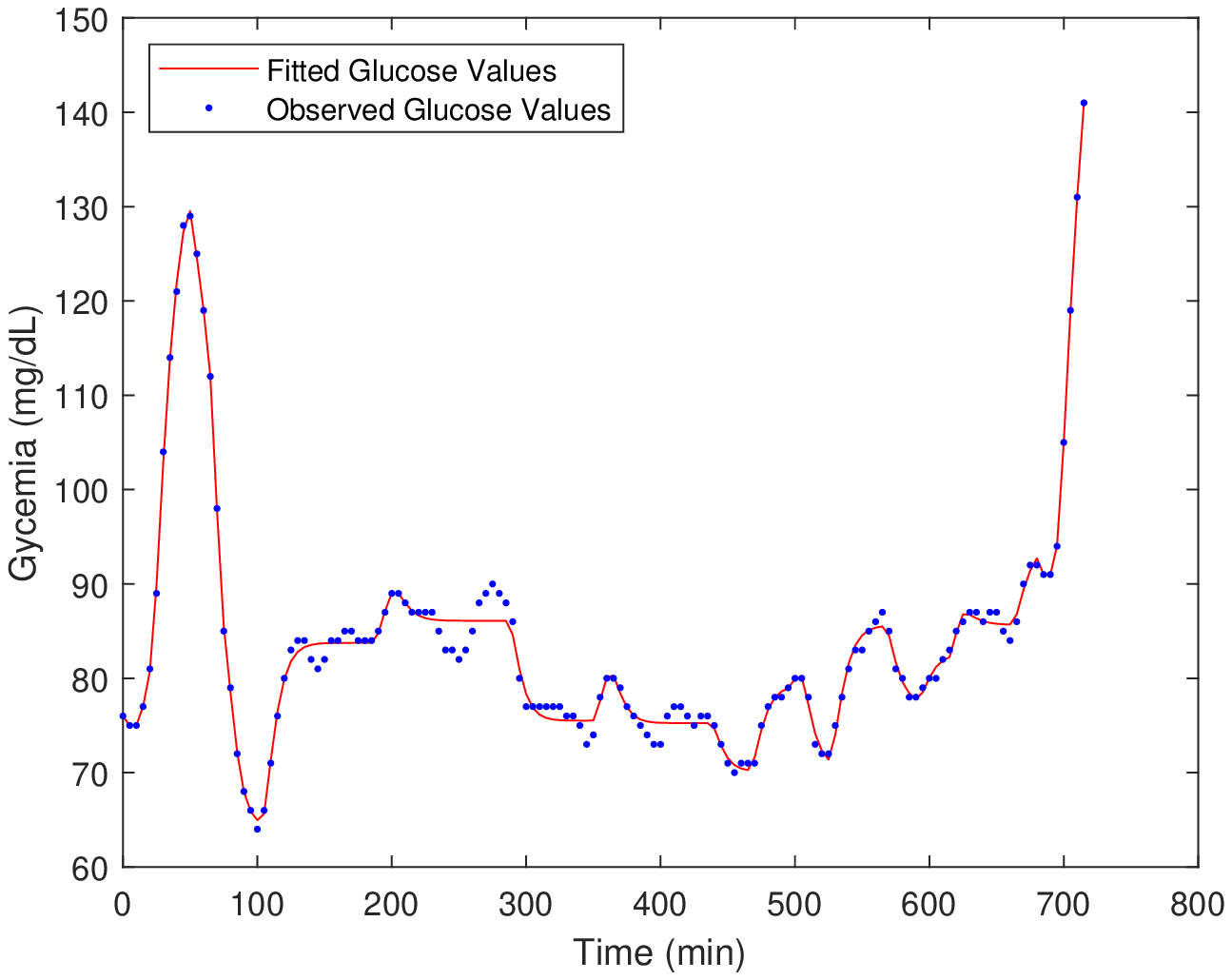}
\subcaption{Gycemia over time, $N_Y=39$.}
\vspace{5.7ex}
\label{MA01b}
   \end{minipage}
\hfill
   \begin{minipage}[b]{0.4\linewidth}
    \includegraphics[width=8cm,height=7cm]{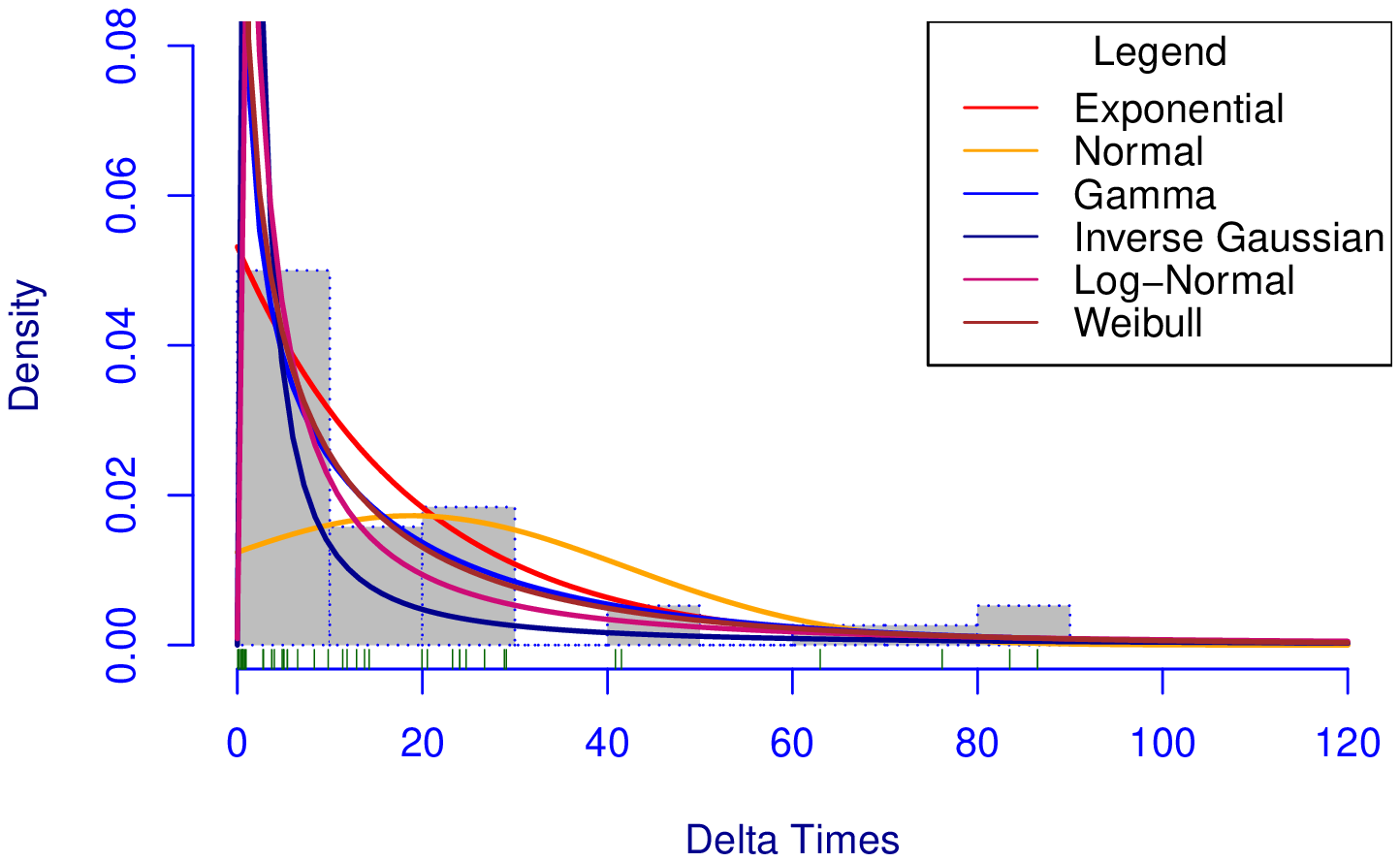}
\subcaption{Histogram of the Delta Times $(\Delta t_i)_{i=1\ldots,N_Y-1}$.}
\vspace{4ex}
\label{MA01c}
   \end{minipage}
  \hfill
   \begin{minipage}[b]{0.4\linewidth}
    \includegraphics[width=8cm,height=7cm]{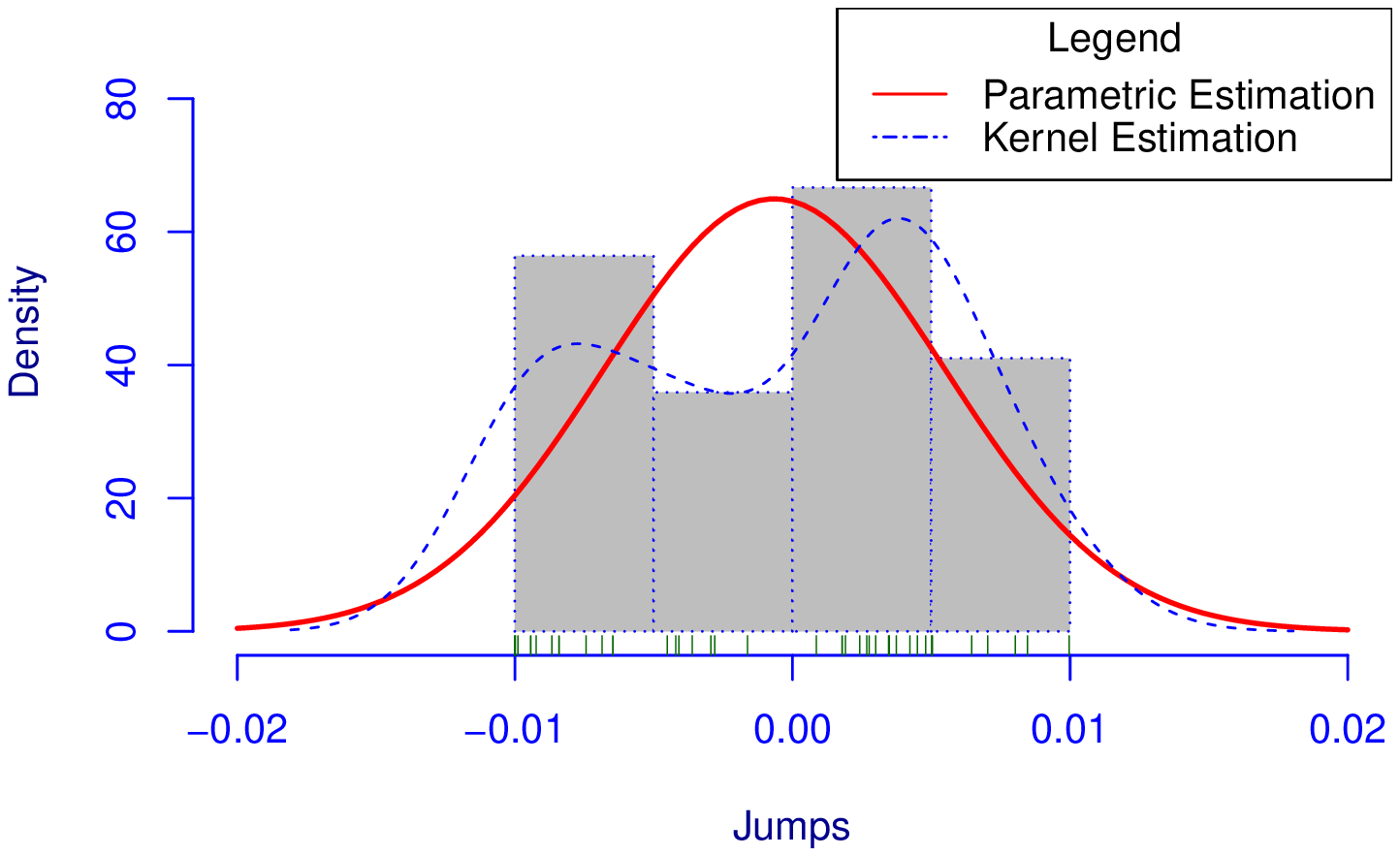}
\subcaption{Estimation of the density of the jumps values.}
\vspace{4ex}
\label{MA01d}
   \end{minipage}
\caption{Results from Patient MA01.}
\label{ResultsSubjectMA01}
\end{figure}

\begin{figure}[H]
  \begin{minipage}[b]{0.4\linewidth}
    \includegraphics[width=8cm, height=6cm]{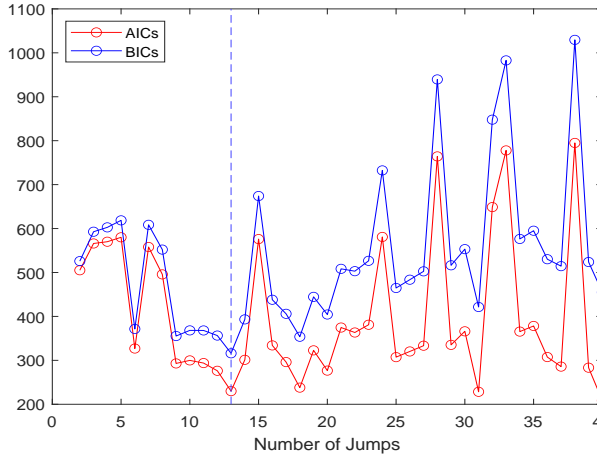}
\subcaption{AIC and BIC trends over the number of jumps.}
\label{PS01a}
\vspace{4ex}
   \end{minipage}
\hfill
   \begin{minipage}[b]{0.4\linewidth}
    \includegraphics[width=8cm,height=6cm]{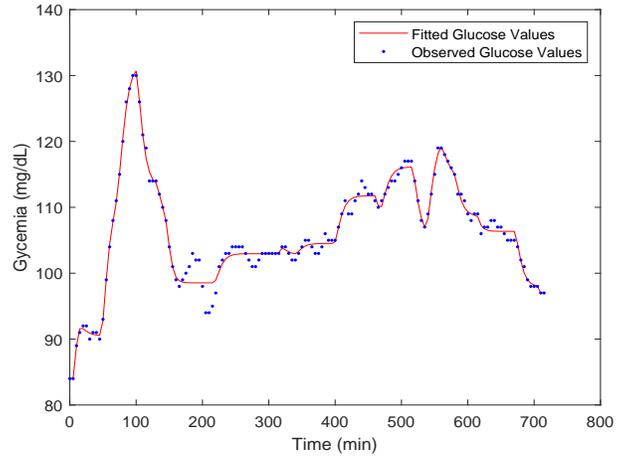}
\subcaption{Gycemia over time, $N_Y=40$.}
\vspace{5.7ex}
\label{PS01b}
   \end{minipage}
\hfill
   \begin{minipage}[b]{0.4\linewidth}
    \includegraphics[width=8cm,height=7cm]{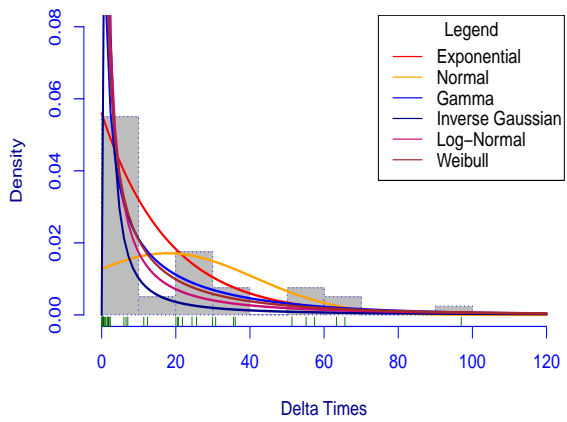}
\subcaption{Histogram of the Delta Times $(\Delta t_i)_{i=1\ldots,N_Y-1}$.}
\vspace{4ex}
\label{PS01c}
   \end{minipage}
  \hfill
   \begin{minipage}[b]{0.4\linewidth}
    \includegraphics[width=8cm,height=7cm]{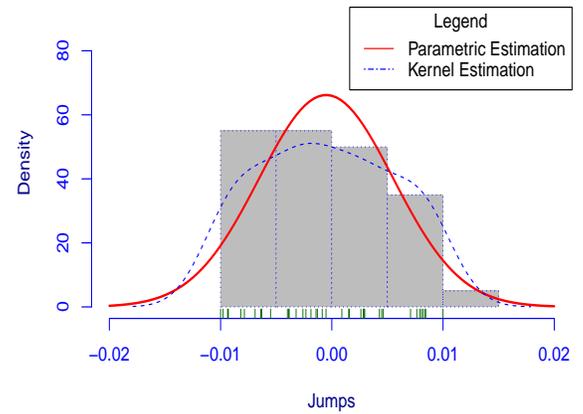}
\subcaption{Estimation of the density of the jumps values.}
\vspace{4ex}
\label{PS01d}
   \end{minipage}
\caption{Results from Patient PS01.}
\label{ResultsSubjectPS01}
\end{figure}

\begin{figure}[H]
  \begin{minipage}[b]{0.4\linewidth}
    \includegraphics[width=8cm, height=6cm]{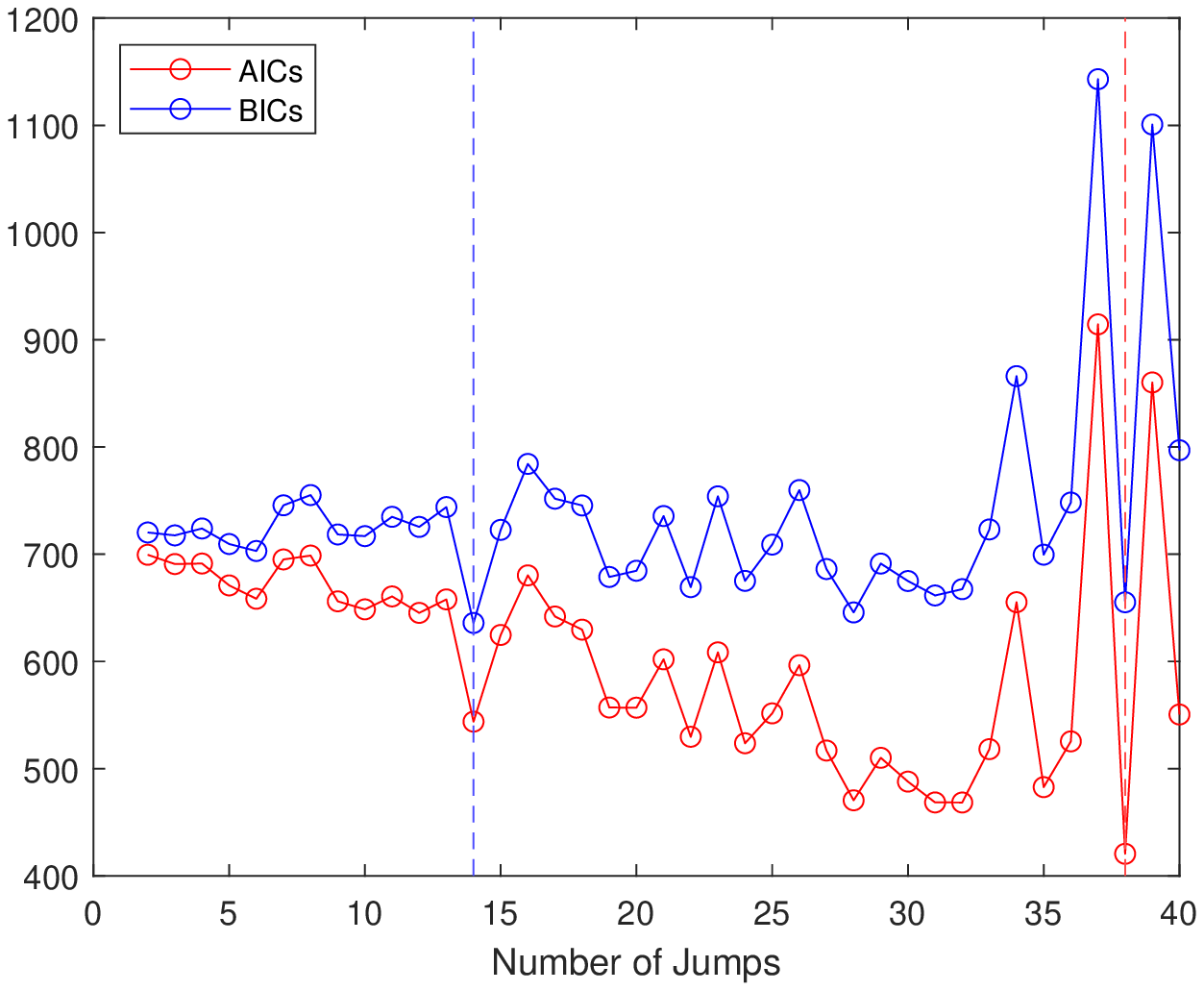}
\subcaption{AIC and BIC trends over the number of jumps.}
\label{TO01a}
\vspace{4ex}
   \end{minipage}
\hfill
   \begin{minipage}[b]{0.4\linewidth}
    \includegraphics[width=8cm,height=6cm]{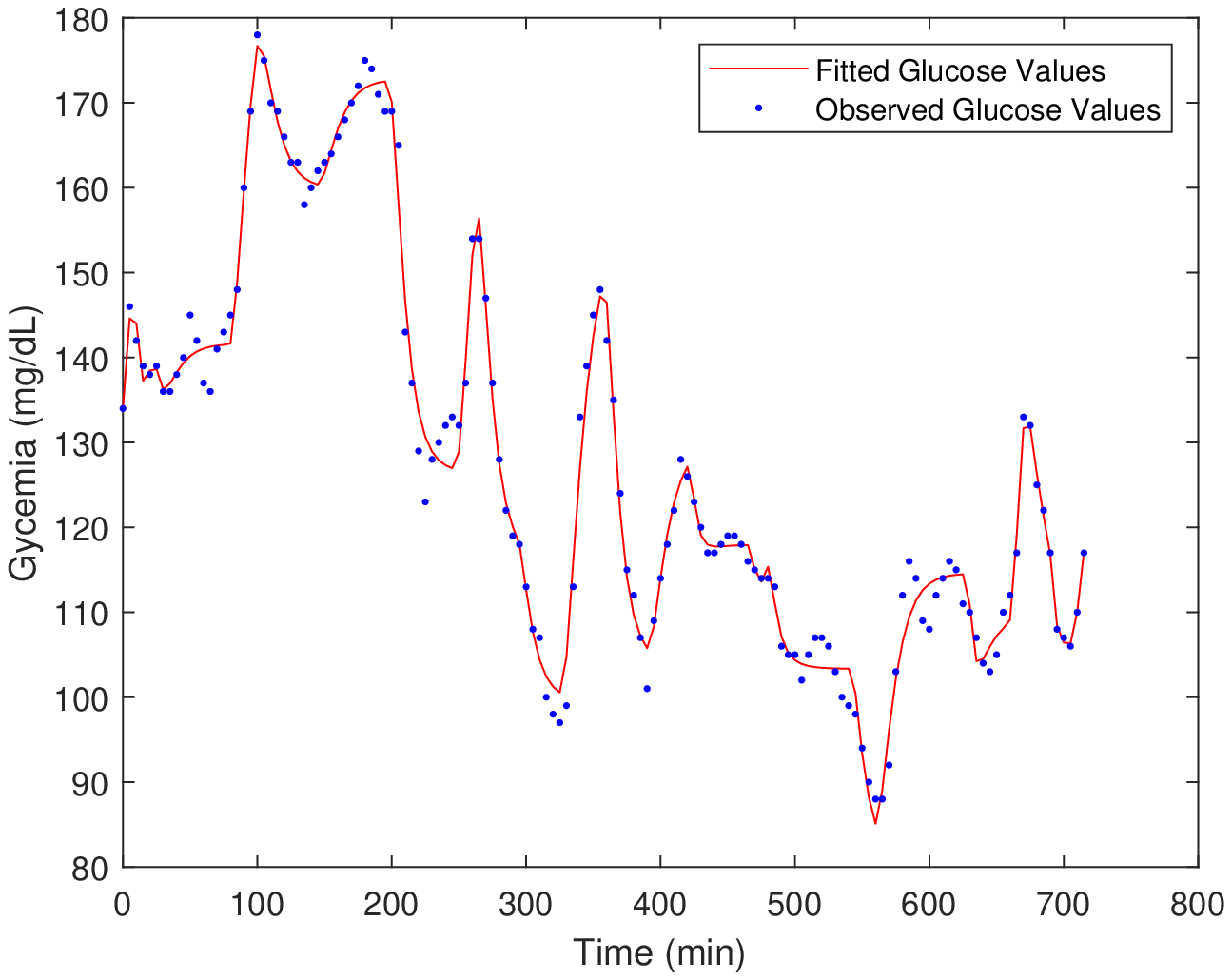}
\subcaption{Gycemia over time, $N_Y=38$.}
\vspace{5.7ex}
\label{TO01b}
   \end{minipage}
\hfill
   \begin{minipage}[b]{0.4\linewidth}
    \includegraphics[width=8cm,height=7cm]{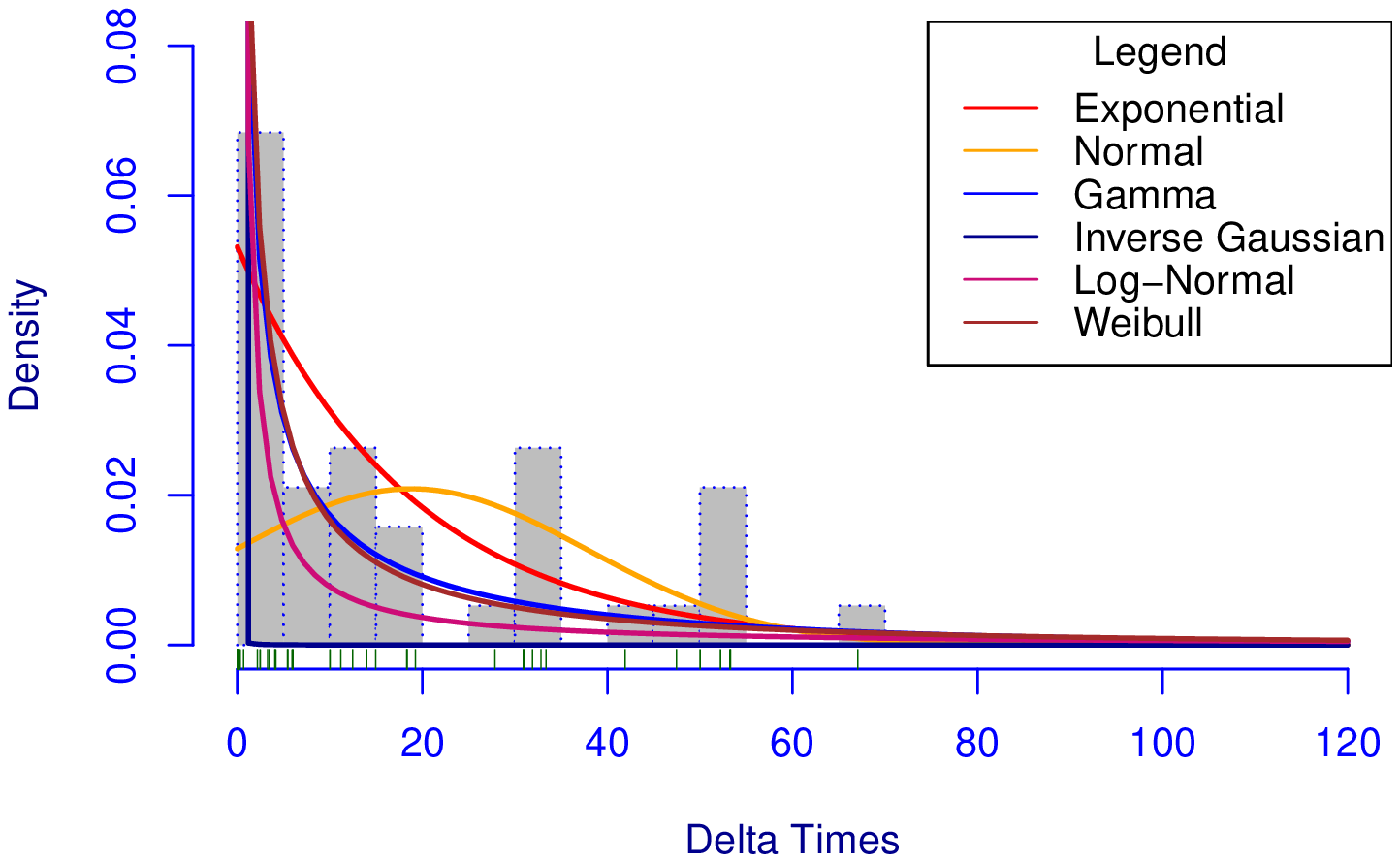}
\subcaption{Histogram of the Delta Times $(\Delta t_i)_{i=1\ldots,N_Y-1}$.}
\vspace{4ex}
\label{TO01c}
   \end{minipage}
  \hfill
   \begin{minipage}[b]{0.4\linewidth}
    \includegraphics[width=8cm,height=7cm]{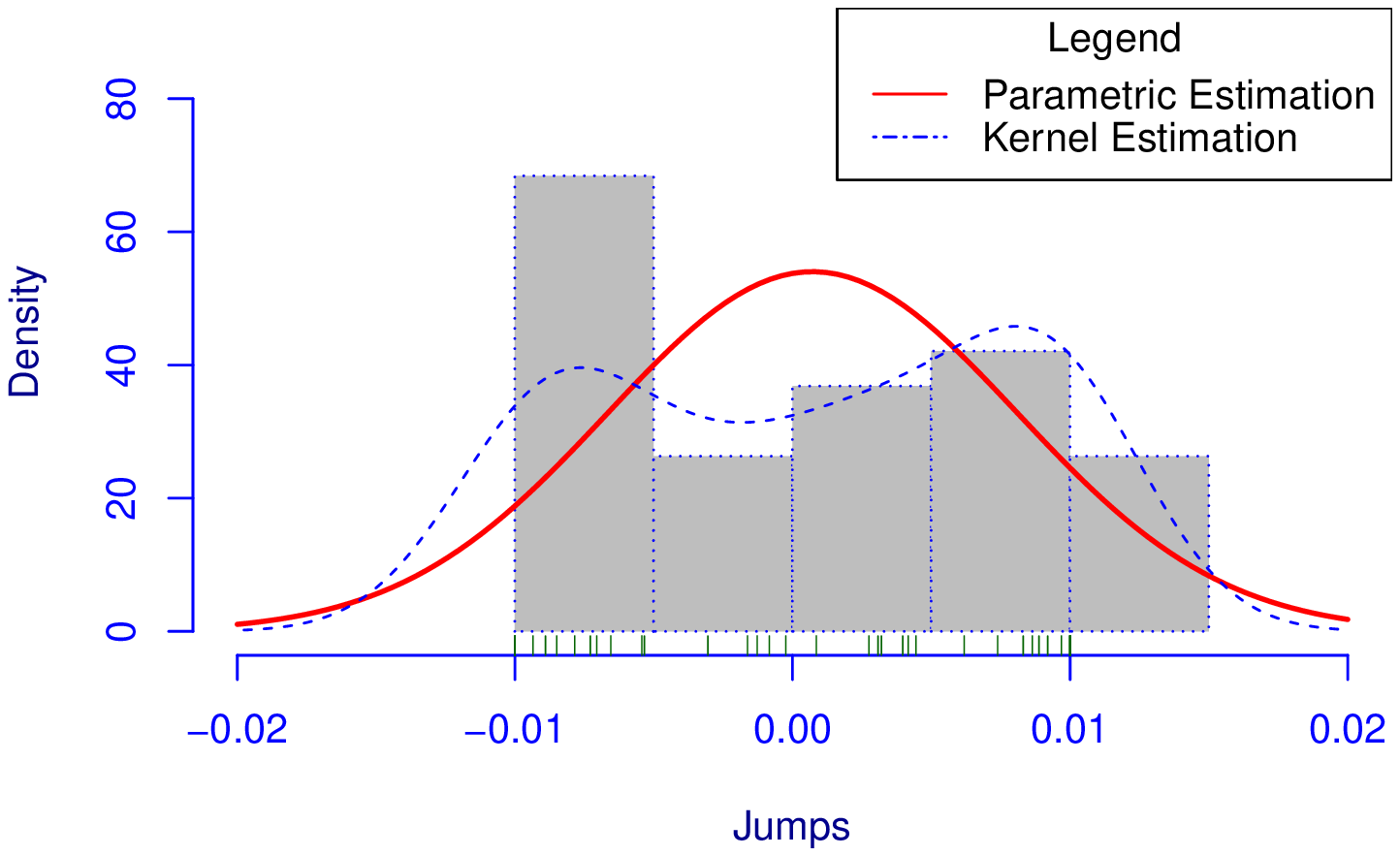}
\subcaption{Estimation of the density of the jumps values.}
\vspace{4ex}
\label{TO01d}
   \end{minipage}
\caption{Results from Patient TO01.}
\label{ResultsSubjectTO01}
\end{figure}

\noindent
In figures \ref{AA01a}, \ref{BO01a}, \ref{CI01a}, \ref{LM01a}, \ref{LR01a}, \ref{MA01a}, \ref{PS01a}, \ref{TO01a}, we show the AIC and BIC trends and the best AIC values of $N_Y$ computed for each patient. As previously mentioned in subsection \ref{Numerical Methods}, the Akaike Information Criterion was preferred and adopted to select the best number of jumps per patient. However, for the sake of completeness, the Bayesian Information Criterion (BIC) values were also calculated. As expected, the Akaike Information Criterion suggests in most cases a model with more parameters to be the best one. Indeed, it can be seen that for the majority of subjects the optimal number of jumps $N_Y$ corresponding to the minimum of the AIC values was notably bigger than the one corresponding to the minimum of the BIC values (dashed red and blue lines of figures \ref{AA01a}, \ref{BO01a}, \ref{CI01a}, \ref{LM01a}, \ref{LR01a}, \ref{MA01a}, \ref{PS01a}, \ref{TO01a}, respectively).

\bigskip

\noindent
As can be seen from figures \ref{AA01b}, \ref{BO01b}, \ref{CI01b}, \ref{LM01b}, \ref{LR01b}, \ref{MA01b},\ref{PS01b}, \ref{TO01b}, the model fits were very good in all cases; with the models corresponding to the minimal AIC values of $N_Y$ being able to capture most of the glucose fluctuations for almost every subject.\\
\bigskip

\noindent
Figures \ref{AA01c}, \ref{BO01c}, \ref{CI01c}, \ref{LM01c}, \ref{LR01c}, \ref{MA01c},\ref{PS01c}, \ref{TO01c} show the histograms of relative frequencies of jump times, $\Delta t_i\ i=1,\ldots, N_Y-1$, where $N_Y$ is the optimal number of jumps for each patient. The classes were calculated with the Sturges method. In addition, on these graphics the Exponential, Normal, Gamma, Inverse Gaussian, Log-Normal and Weibull distributions fitted using the R gamlss package~\cite{gamlss} were superimposed.

\bigskip

\noindent
The frequency histograms associated with the intensity of the jumps $(Y_i)_{i=1\ldots N_Y}$ are shown in figures \ref{AA01d}, \ref{BO01d}, \ref{CI01d}, \ref{LM01d}, \ref{LR01d}, \ref{MA01d}, \ref{PS01d}, \ref{TO01d}. In these panels, estimated density distributions obtained through both parametric and non-parametric methods have been superimposed. Unlike the non-parametric Kernel method~\cite{Silverman}, which makes a mixture of Gaussians generated in correspondence with each statistical unit, the parametric estimation corresponds to a single distribution which we supposed to be a standard normal one. Table \ref{tableall} shows for each patient the means and the standard deviations extracted from them. As expected, the first method delivers a closer fit to the histograms.

%%%%%%%%%%%%
\subsubsection*{Patient AA01}

%%%%%%%%%%%%
\subsubsection*{Patient BO01}

%%%%%%%%%%%%
\subsubsection*{Patient CI01}

%%%%%%%%%%%%
\subsubsection*{Patient LM01}

%%%%%%%%%%%%
\subsubsection*{Patient LR01}

%%%%%%%%%%%%
\subsubsection*{Patient MA01}

%%%%%%%%%%%%
\subsubsection*{Patient PS01}

%%%%%%%%%%%%
\subsubsection*{Patient TO01}

\begin{table}[H]
\centering
\begin{tabular}{||l|l|l||}
  \hline
\textbf{Patient}& \textbf{Expectation $\mu$} & \textbf{Standard Deviation $\sigma$} \\
  \hline
AA01 & -0.0001626885 & 0.006963786 \\
\hline
BO01&  -0.0000212980 &  0.006082102\\
\hline
CI01 & -0.0003323163 & 0.005998925\\
\hline
LM01 & 0.00007391962 & 0.006144557 \\
\hline
LR01 & -0.0006154387 & 0.007353402\\
\hline
MA01& -0.0006552171 & 0.006140687 \\
\hline
PS01&  -0.0005062152 & 0.00602809 \\
\hline
TO01 & 0.0007230374 & 0.007385183 \\
   \hline
\end{tabular}
\caption{Means and the standard deviations obtained from the normal parametric density for each patient.}
\label{tableall}
\end{table}

\noindent
In order to assess which probability distribution was most suited to fit the waiting times, An analysis of the goodness of the fittings was performed. Specifically, a comparison between the Exponential, Gamma, Normal, Log-Normal, Weibull and Inverse Gaussian distributions was carried out.
%DA CANCELLAREWhile we may speculate that the relatively poorer adaptation of the exponential distribution to the waiting times might be due to the small number of jumps identified in a 12 hours period, the parameter $\lambda$ thus estimated may be taken as a first approximation to the assessment of the subject's profile insofar as frequency of perturbations is concerned: the higher the value of $\lambda$ the lower the average waiting time between jumps and the higher the frequency of (clinically significant) perturbations.

\begin{table}[H]
\centering
\begin{tabular}{||l|l||}
  \hline
 \textbf{Distribution} &  \textbf{AIC }\\
  \hline
Exponential  & 244.62 \\
  Gamma  & 245.25 \\
Weibull & 245.77 \\
  Log-Normal & 258.20 \\
  Normal & 269.12 \\
 Inverse-Gaussian & 327.99 \\
     \hline
\end{tabular}
\caption{Goodness of fit $\Delta t_i$ Patient AA01}
\label{AA01table}
\end{table}

\begin{table}[H]
\centering
\begin{tabular}{||l|l||}
  \hline
 \textbf{Distribution} &  \textbf{AIC }\\
  \hline
Gamma & 272.7566 \\
  Weibull & 275.8100 \\
Log-Normal  &292.0124\\
  Exponential & 301.0369\\
    Normal & 359.7758 \\
  Inverse-Gaussian & 474.6699\\
     \hline
\end{tabular}
\caption{Goodness of fit $\Delta t_i$ Patient B001}
\label{BO01table}
\end{table}

\begin{table}[H]
\centering
\begin{tabular}{||l|l||}
  \hline
\textbf{Distribution} &  \textbf{AIC }\\
  \hline
  Gamma & 162.8705 \\
  Weibull & 172.5374 \\
Log-Normal  & 206.9017\\
  Exponential & 219.6703 \\
    Normal & 249.8459 \\
  Inverse-Gaussian & 922.6549\\
   \hline
\end{tabular}
\caption{Goodness of fit $\Delta t_i$ Patient CI01}
\label{CI01table}
\end{table}

\begin{table}[H]
\centering
\begin{tabular}{||l|l||}
  \hline
\textbf{Distribution} &  \textbf{AIC }\\
  \hline
Gamma & 226.6330 \\
  Weibull & 231.8983 \\
  Exponential & 245.8892 \\
  Log-Normal  &248.5496\\
    Normal & 273.8758 \\
  Inverse-Gaussian & 399.9703\\
   \hline
\end{tabular}
\caption{Goodness of fit $\Delta t_i$ Patient LM01}
\label{LM01table}
\end{table}

\begin{table}[H]
\centering
\begin{tabular}{||l|l||}
  \hline
\textbf{Distribution} &  \textbf{AIC }\\
  \hline
    Gamma & 239.1154 \\
  Weibull & 244.0272 \\
  Exponential & 258.5743 \\
Log-Normal  &262.3403\\
      Normal & 296.7596 \\
  Inverse-Gaussian & 410.8386\\
   \hline
\end{tabular}
\caption{Goodness of fit $\Delta t_i$ Patient LR01}
\label{LR01table}
\end{table}

\begin{table}[H]
\centering
\begin{tabular}{||l|l||}
  \hline
\textbf{Distribution} &  \textbf{AIC }\\
  \hline
 Gamma & 295.4941 \\
  Weibull & 295.7222\\
  Exponential & 301.0369 \\
Log-Normal  &301.4513\\
Inverse-Gaussian & 326.3976\\
      Normal & 350.4726 \\
     \hline
\end{tabular}
\caption{Goodness of fit $\Delta t_i$ Patient MA01}
\label{MA01table}
\end{table}

\begin{table}[H]
\centering
\begin{tabular}{||l|l||}
  \hline
 \textbf{Distribution} &  \textbf{AIC }\\
  \hline
  Gamma & 289.1986 \\
  Weibull & 289.5733\\
Log-Normal  &291.2840\\
 Inverse-Gaussian & 305.0637\\
  Exponential & 312.6722\\
    Normal & 369.4195 \\
      \hline
\end{tabular}
\caption{Goodness of fit $\Delta t_i$ Patient PS01}
\label{PS01table}
\end{table}

\begin{table}[H]
\centering
\begin{tabular}{||l|l||}
  \hline
\textbf{Distribution} &  \textbf{AIC }\\
  \hline
Gamma & 249.02 \\
  Weibull & 261.46 \\
Log-Normal  &296.82\\
  Exponential & 301.04 \\
    Normal & 336.10 \\
  Inverse-Gaussian & 741.39\\
   \hline
\end{tabular}
\caption{Goodness of fit $\Delta t_i$ Patient TO01}
\label{TO01table}
\end{table}

\noindent
According to Tables \ref{AA01table}, \ref{BO01table}, \ref{CI01table}, \ref{LM01table}, \ref{LR01table}, \ref{MA01table}, \ref{PS01table}, \ref{TO01table}, the Gamma distribution appears to be the optimal one. We report in table \ref{GammaParameters}, with the gamlss.family notation ~\cite{gamlss}, the corresponding estimated parameters through the Maximum Likelihood Estimation (MLE) method.

\begin{table}[H]
\centering
\begin{tabular}{||l|l|l|l||}
  \hline
\textbf{Patient}& $N_Y$ & \textbf{$\mu$} & \textbf{$\sigma$} \\
  \hline
AA01 &29 & 24.12264 & 1.136236  \\
\hline
BO01& 39& 18.81577&  1.563323\\
\hline
CI01 & 26&28.60001& 1.971192\\
\hline
LM01 &29& 24.65488 & 1.540134 \\
\hline
LR01 & 31& 23.06438& 1.524811\\
\hline
MA01&  39&18.81577& 1.279943\\
\hline
PS01& 40&17.87475  & 1.503246 \\
\hline
TO01 & 38& 18.81579& 1.753733  \\
   \hline
\end{tabular}
\caption{Parameters of the Gamma Distribution estimated for each patient.}
\label{GammaParameters}
\end{table}

\noindent
For the sake of completeness, the best fit of parameter values for each subject (according to the Akaike Information Criterion) is shown in tables \ref{kinetics}, \ref{saltietempiAA01}, \ref{saltietempiBO01}, \ref{saltietempiCI01}, \ref{saltietempiLM01}, \ref{saltietempiLR01}, \ref{saltietempiMA01},\ref{saltietempiPS01}, \ref{saltietempiTO01}. \\

\begin{center}
\tablefirsthead{%
\hline
\multicolumn{1}{||l|}{\textbf{Patient}} &
\multicolumn{1}{|l|}{$\mathbf{N_Y}$} &
\multicolumn{1}{|c|}{$\mathbf{k_{EH}}$} &
\multicolumn{1}{|c|}{$\mathbf{H_0}$} &
\multicolumn{1}{|c||}{$\mathbf{Y_0}$} \\
\hline}
\tablehead{%
\hline
\multicolumn{5}{|l|}{\small\sl continued from previous page}\\
\hline
\multicolumn{1}{||l|}{\textbf{Patient}} &
\multicolumn{1}{|l|}{$\mathbf{N_Y}$} &
\multicolumn{1}{|l|}{$\mathbf{k_{EH}}$} &
\multicolumn{1}{|l|}{$\mathbf{H_0}$} &
\multicolumn{1}{|l||}{$\mathbf{Y_0}$} \\
\hline}
\tabletail{%
\hline
\multicolumn{5}{|l|}{\small\sl continued on next page}\\
\hline}
\tablelasttail{\hline}
\bottomcaption{Kinetic parameters per patient according to the best number of Jumps $N_Y$.}
\begin{supertabular}[H]{||l|l|l|l|l||}
AA01 & 29 &1.5151577e-01 &  1.9999999e-01 & -9.9972279e-01 \\
\hline
BO01 & 39  &1.9999963e-01 &  1.9999985e-01  & 9.9999921e-01 \\
\hline
CI01& 26 &1.9999999e-01 &   2.0909944e-02  & 6.9208498e-01 \\
\hline
LM01 & 29 &1.9999996e-01  & 1.5973173e-01 & -9.9999911e-01\\
\hline
LR01& 31 &1.6221208e-01 &  1.4979140e-01 &  -9.4417929e-01\\
\hline
MA01& 39 &1.9999997e-01   &1.8175307e-01 & -9.0751673e-02 \\
\hline
PS01 & 40 &1.9414967e-01  & 2.0000000e-01 & -8.7246572e-01 \\
\hline
TO01 & 38 &1.0272332e-01  & 2.0000000e-01 & -9.9980120e-01 \\
\hline
\end{supertabular}
\label{kinetics}
\end{center}

\newpage

\begin{table}[!ht]
    \centering
    \begin{tabular}{||l|l|l|l|l||}
    \hline
        $\mathbf{t_1}$ &$\mathbf{t_2}$ & $\mathbf{t_3}$ & $\mathbf{t_4}$ & $\mathbf{t_5}$ \\ \hline
        5.2364027 & 12.8370142 & 41.5858319 & 77.7234427 & 80.6372994 \\ \hline
        $\mathbf{Y_1}$ & $\mathbf{Y_2}$ & $\mathbf{Y_3}$ & $\mathbf{Y_4}$ & $\mathbf{Y_5}$\\ \hline
        -0.0031786 & -0.0099723 & 0.0078998 & -0.0054477 & 0.0100000 \\ \hline
       $\mathbf{t_6}$ & $\mathbf{t_7}$ & $\mathbf{t_8}$ & $\mathbf{t_9}$ & $\mathbf{t_{10}}$ \\ \hline
        86.7001226 & 86.7120607 & 175.3760136 & 181.0414045 & 194.3757596 \\ \hline
        $\mathbf{Y_6}$ & $\mathbf{Y_7}$ & $\mathbf{Y_8}$ & $\mathbf{Y_9}$ & $\mathbf{Y_{10}}$ \\ \hline
        0.0046722 & -0.0100000 & 0.0032524 & -0.0075409 & 0.0044350 \\ \hline
       $\mathbf{t_{11}}$ & $\mathbf{t_{12}}$ & $\mathbf{t_{13}}$ & $\mathbf{t_{14}}$ & $\mathbf{t_{15}}$ \\ \hline
        232.3488651 & 236.7362091 & 277.0105840 & 335.8868838 & 338.5933877 \\ \hline
         $\mathbf{Y_{11}}$ & $\mathbf{Y_{12}}$ & $\mathbf{Y_{13}}$ & $\mathbf{Y_{14}}$ & $\mathbf{Y_{15}}$ \\ \hline
        0.0071978 & -0.0043805 & -0.0017729 & 0.0060891 & -0.0044720 \\ \hline
        $\mathbf{t_{16}}$ & $\mathbf{t_{17}}$ & $\mathbf{t_{18}}$ & $\mathbf{t_{19}}$ & $\mathbf{t_{20}}$  \\ \hline
        364.7587451 & 370.7980945 & 443.1351053 & 445.5294229 & 505.5644604 \\ \hline
        $\mathbf{Y_{16}}$ & $\mathbf{Y_{17}}$ & $\mathbf{Y_{18}}$ & $\mathbf{Y_{19}}$ & $\mathbf{Y_{20}}$ \\ \hline
        -0.0073427 & 0.0089861 & 0.0069922 & -0.0088973 & 0.0064212 \\ \hline
        $\mathbf{t_{21}}$ & $\mathbf{t_{22}}$ & $\mathbf{t_{23}}$ & $\mathbf{t_{24}}$ & $\mathbf{t_{25}}$ \\ \hline
        520.5688299 & 545.8138547 & 546.7999130 & 560.8694650 & 568.7932271 \\ \hline
        $\mathbf{Y_{21}}$ & $\mathbf{Y_{22}}$ & $\mathbf{Y_{23}}$ & $\mathbf{Y_{24}}$ & $\mathbf{Y_{25}}$\\ \hline
        -0.0042245 & -0.0100000 & 0.0083088 & -0.0065152 & 0.0041529 \\ \hline
        $\mathbf{t_{26}}$ & $\mathbf{t_{27}}$ & $\mathbf{t_{28}}$ & $\mathbf{t_{29}}$ & \\ \hline
        578.0625743 & 626.2670384 & 668.4789765 & 699.5566696 & ~ \\ \hline
        $\mathbf{Y_{26}}$ & $\mathbf{Y_{27}}$ & $\mathbf{Y_{28}}$ & $\mathbf{Y_{29}}$ &   \\ \hline
        0.0030111 & -0.0023915 & -0.0100000 & 0.0100000 & \\ \hline
    \end{tabular}
    \caption{Patient AA01: Jump Times and Intensities of the Jumps , $N_Y$=29.}
    \label{saltietempiAA01}
\end{table}

\newpage
\begin{table}[!ht]
    \centering
    \begin{tabular}{||l|l|l|l|l||}
    \hline
        $\mathbf{t_1}$ &$\mathbf{t_2}$ & $\mathbf{t_3}$ & $\mathbf{t_4}$ & $\mathbf{t_5}$ \\ \hline
        1.7531642 & 6.9422950 & 9.6426214 & 14.9446212 & 15.2104916 \\ \hline
        $\mathbf{Y_1}$ & $\mathbf{Y_2}$ & $\mathbf{Y_3}$ & $\mathbf{Y_4}$ & $\mathbf{Y_5}$\\ \hline
        -0.0077292 & 0.0099049 & -0.0087262 & 0.0033857 & 0.0019620 \\ \hline
       $\mathbf{t_6}$ & $\mathbf{t_7}$ & $\mathbf{t_8}$ & $\mathbf{t_9}$ & $\mathbf{t_{10}}$ \\ \hline
        20.8367196 & 21.6261287 & 56.1390841 & 81.2205516 & 91.3284842 \\ \hline
       $\mathbf{Y_6}$ & $\mathbf{Y_7}$ & $\mathbf{Y_8}$ & $\mathbf{Y_9}$ & $\mathbf{Y_{10}}$ \\ \hline
        -0.0014224 & -0.0009051 & -0.0017687 & -0.0023248 & 0.0046516 \\ \hline
        $\mathbf{t_{11}}$ & $\mathbf{t_{12}}$ & $\mathbf{t_{13}}$ & $\mathbf{t_{14}}$ & $\mathbf{t_{15}}$ \\ \hline
        120.9721043 & 121.8619375 & 132.0787944 & 150.5280246 & 150.7901859 \\ \hline
         $\mathbf{Y_{11}}$ & $\mathbf{Y_{12}}$ & $\mathbf{Y_{13}}$ & $\mathbf{Y_{14}}$ & $\mathbf{Y_{15}}$ \\ \hline
        -0.0079149 & 0.0042477 & -0.0030421 & -0.0034532 & 0.0098593 \\ \hline
       $\mathbf{t_{16}}$ & $\mathbf{t_{17}}$ & $\mathbf{t_{18}}$ & $\mathbf{t_{19}}$ & $\mathbf{t_{20}}$  \\ \hline
        165.8442276 & 179.0751796 & 242.4253110 & 243.5121682 & 273.0707517 \\ \hline
        $\mathbf{Y_{16}}$ & $\mathbf{Y_{17}}$ & $\mathbf{Y_{18}}$ & $\mathbf{Y_{19}}$ & $\mathbf{Y_{20}}$ \\ \hline
        -0.0041980 & 0.0037630 & 0.0043378 & -0.0052634 & -0.0071043 \\ \hline
       $\mathbf{t_{21}}$ & $\mathbf{t_{22}}$ & $\mathbf{t_{23}}$ & $\mathbf{t_{24}}$ & $\mathbf{t_{25}}$ \\ \hline
        273.3078462 & 332.0561503 & 405.1572658 & 405.2791883 & 420.2552715 \\ \hline
       $\mathbf{Y_{21}}$ & $\mathbf{Y_{22}}$ & $\mathbf{Y_{23}}$ & $\mathbf{Y_{24}}$ & $\mathbf{Y_{25}}$\\ \hline
        0.0060889 & 0.0013333 & 0.0020822 & 0.0032244 & -0.0069572 \\ \hline
       $\mathbf{t_{26}}$ & $\mathbf{t_{27}}$ & $\mathbf{t_{28}}$ & $\mathbf{t_{29}}$ & \\ \hline
        437.9785828 & 567.8344144 & 585.2840801 & 626.8542070 & 629.2681306 \\ \hline
        $\mathbf{Y_{26}}$ & $\mathbf{Y_{27}}$ & $\mathbf{Y_{28}}$ & $\mathbf{Y_{29}}$ & $\mathbf{Y_{30}}$   \\ \hline
        0.0018411 & 0.0029846 & -0.0024690 & -0.0067185 & 0.0066140 \\ \hline
        $\mathbf{t_{31}}$ & $\mathbf{t_{32}}$ & $\mathbf{t_{33}}$ & $\mathbf{t_{34}}$ & $\mathbf{t_{35}}$ \\ \hline
        643.2776360 & 643.9756707 & 675.6225995 & 711.4105311 & 711.6292886 \\ \hline
        $\mathbf{Y_{31}}$ & $\mathbf{Y_{32}}$ & $\mathbf{Y_{33}}$ & $\mathbf{Y_{34}}$ & $\mathbf{Y_{35}}$ \\ \hline
        0.0097737 & -0.0100000 & -0.0017152 & -0.0082753 & 0.0066267 \\ \hline
        $\mathbf{t_{36}}$ & $\mathbf{t_{37}}$ & $\mathbf{t_{38}}$ & $\mathbf{t_{39}}$ & \\ \hline
        714.9978196 & 714.9998922 & 715.0000000 & 715.0000000 & ~ \\ \hline
        $\mathbf{Y_{36}}$ & $\mathbf{Y_{37}}$ & $\mathbf{Y_{38}}$ & $\mathbf{Y_{39}}$ &    \\ \hline
        -0.0035245 & 0.0100000 & -0.0099993 & 0.0099996 &\\ \hline
    \end{tabular}
    \caption{Patient BO01: Jump Times and Intensities of the Jumps , $N_Y$=39.}
    \label{saltietempiBO01}
\end{table}

\newpage
\begin{table}[!ht]
    \centering
    \begin{tabular}{||l|l|l|l|l||}
    \hline
       $\mathbf{t_1}$ &$\mathbf{t_2}$ & $\mathbf{t_3}$ & $\mathbf{t_4}$ & $\mathbf{t_5}$ \\ \hline
        0.0000000 & 0.0000000 & 1.7661084 & 2.4728315 & 23.8122309 \\ \hline
        $\mathbf{Y_1}$ & $\mathbf{Y_2}$ & $\mathbf{Y_3}$ & $\mathbf{Y_4}$ & $\mathbf{Y_5}$\\ \hline
        -0.0040950 & -0.0068447 & 0.0074244 & 0.0034137 & -0.0009755 \\ \hline
        $\mathbf{t_6}$ & $\mathbf{t_7}$ & $\mathbf{t_8}$ & $\mathbf{t_9}$ & $\mathbf{t_{10}}$ \\ \hline
        62.9555029 & 105.7246465 & 141.1139283 & 180.2594550 & 243.0826317 \\ \hline
        $\mathbf{Y_6}$ & $\mathbf{Y_7}$ & $\mathbf{Y_8}$ & $\mathbf{Y_9}$ & $\mathbf{Y_{10}}$ \\ \hline
        -0.0003314 & 0.0020335 & -0.0025940 & 0.0025657 & -0.0008888 \\ \hline
        $\mathbf{t_{11}}$ & $\mathbf{t_{12}}$ & $\mathbf{t_{13}}$ & $\mathbf{t_{14}}$ & $\mathbf{t_{15}}$ \\ \hline
        319.5023530 & 442.2855711 & 444.3017886 & 457.5395079 & 499.8336690 \\ \hline
       $\mathbf{Y_{11}}$ & $\mathbf{Y_{12}}$ & $\mathbf{Y_{13}}$ & $\mathbf{Y_{14}}$ & $\mathbf{Y_{15}}$ \\ \hline
        -0.0001726 & 0.0047068 & -0.0051530 & 0.0009582 & -0.0091218 \\ \hline
        $\mathbf{t_{16}}$ & $\mathbf{t_{17}}$ & $\mathbf{t_{18}}$ & $\mathbf{t_{19}}$ & $\mathbf{t_{20}}$  \\ \hline
        501.1904024 & 584.7252873 & 584.8950891 & 665.1868158 & 668.4435777 \\ \hline
        $\mathbf{Y_{16}}$ & $\mathbf{Y_{17}}$ & $\mathbf{Y_{18}}$ & $\mathbf{Y_{19}}$ & $\mathbf{Y_{20}}$ \\ \hline
        0.0086406 & -0.0091392 & 0.0094690 & -0.0059866 & 0.0067016 \\ \hline
       $\mathbf{t_{21}}$ & $\mathbf{t_{22}}$ & $\mathbf{t_{23}}$ & $\mathbf{t_{24}}$ & $\mathbf{t_{25}}$ \\ \hline
        695.5277727 & 698.3367440 & 705.1897997 & 714.8207471 & 714.9125837 \\ \hline
         $\mathbf{Y_{21}}$ & $\mathbf{Y_{22}}$ & $\mathbf{Y_{23}}$ & $\mathbf{Y_{24}}$ & $\mathbf{Y_{25}}$\\ \hline
        -0.0062823 & 0.0063404 & -0.0031597 & -0.0100000 & 0.0099975 \\ \hline
        $\mathbf{t_{26}}$ & ~ & ~ & ~ & ~ \\ \hline
        714.9993925 & ~ & ~ & ~ & ~ \\ \hline
        $\mathbf{Y_{26}}$ & ~ & ~ & ~ & ~ \\ \hline
        -0.0061470 & ~ & ~ & ~ & ~\\ \hline
    \end{tabular}
    \caption{Patient CI01: Jump Times and Intensities of the Jumps , $N_Y$=26.}
    \label{saltietempiCI01}
\end{table}

\newpage
\begin{table}[!ht]
    \centering
    \begin{tabular}{||l|l|l|l|l||}
    \hline
         $\mathbf{t_1}$ &$\mathbf{t_2}$ & $\mathbf{t_3}$ & $\mathbf{t_4}$ & $\mathbf{t_5}$ \\ \hline
        36.6884378 & 36.7057686 & 67.9336347 & 110.8990587 & 180.4096177 \\ \hline
        $\mathbf{Y_1}$ & $\mathbf{Y_2}$ & $\mathbf{Y_3}$ & $\mathbf{Y_4}$ & $\mathbf{Y_5}$\\ \hline
        0.0051069 & -0.0023904 & -0.0036958 & -0.0024041 & -0.0044771 \\ \hline
       $\mathbf{t_6}$ & $\mathbf{t_7}$ & $\mathbf{t_8}$ & $\mathbf{t_9}$ & $\mathbf{t_{10}}$ \\ \hline
        183.9066619 & 209.1660485 & 240.9189356 & 256.0984147 & 302.4857421 \\ \hline
        $\mathbf{Y_6}$ & $\mathbf{Y_7}$ & $\mathbf{Y_8}$ & $\mathbf{Y_9}$ & $\mathbf{Y_{10}}$ \\ \hline
        0.0061199 & -0.0020663 & 0.0064213 & -0.0038317 & -0.0013097 \\ \hline
        $\mathbf{t_{11}}$ & $\mathbf{t_{12}}$ & $\mathbf{t_{13}}$ & $\mathbf{t_{14}}$ & $\mathbf{t_{15}}$ \\ \hline
        325.7139632 & 341.3859477 & 342.5575031 & 345.2138864 & 345.2910954 \\ \hline
        $\mathbf{Y_{11}}$ & $\mathbf{Y_{12}}$ & $\mathbf{Y_{13}}$ & $\mathbf{Y_{14}}$ & $\mathbf{Y_{15}}$ \\ \hline
        0.0022173 & 0.0099747 & -0.0060372 & -0.0044706 & 0.0012415 \\ \hline
         $\mathbf{t_{16}}$ & $\mathbf{t_{17}}$ & $\mathbf{t_{18}}$ & $\mathbf{t_{19}}$ & $\mathbf{t_{20}}$  \\ \hline
        396.8162060 & 415.4458159 & 432.3201529 & 544.0285164 & 544.0408002 \\ \hline
        $\mathbf{Y_{16}}$ & $\mathbf{Y_{17}}$ & $\mathbf{Y_{18}}$ & $\mathbf{Y_{19}}$ & $\mathbf{Y_{20}}$ \\ \hline
        0.0042504 & 0.0082136 & -0.0100000 & -0.0077589 & 0.0043591 \\ \hline
       $\mathbf{t_{21}}$ & $\mathbf{t_{22}}$ & $\mathbf{t_{23}}$ & $\mathbf{t_{24}}$ & $\mathbf{t_{25}}$ \\ \hline
        574.7776621 & 594.4572234 & 644.5672747 & 696.0315707 & 698.2948359 \\ \hline
       $\mathbf{Y_{21}}$ & $\mathbf{Y_{22}}$ & $\mathbf{Y_{23}}$ & $\mathbf{Y_{24}}$ & $\mathbf{Y_{25}}$\\ \hline
        0.0062931 & -0.0041780 & 0.0026677 & -0.0051616 & 0.0029556 \\ \hline
        $\mathbf{t_{26}}$ & $\mathbf{t_{27}}$ & $\mathbf{t_{28}}$ & $\mathbf{t_{29}}$ &  \\ \hline
        712.9636084 & 714.6250258 & 714.9998962 & 714.9999976 & ~ \\ \hline
        $\mathbf{Y_{26}}$ & $\mathbf{Y_{27}}$ & $\mathbf{Y_{28}}$ & $\mathbf{Y_{29}}$ &  \\ \hline
        0.0100000 & -0.0099142 & 0.0099959 & -0.0099779 &\\ \hline
    \end{tabular}
    \caption{Patient LM01: Jump Times and Intensities of the Jumps , $N_Y$=29.}
    \label{saltietempiLM01}
\end{table}

\newpage

\begin{table}[!ht]
    \centering
    \begin{tabular}{||l|l|l|l|l||}
    \hline
        $\mathbf{t_1}$ &$\mathbf{t_2}$ & $\mathbf{t_3}$ & $\mathbf{t_4}$ & $\mathbf{t_5}$ \\ \hline
        11.9094990 & 12.1019352 & 18.8857897 & 60.1995371 & 80.3116479 \\ \hline
       $\mathbf{Y_1}$ & $\mathbf{Y_2}$ & $\mathbf{Y_3}$ & $\mathbf{Y_4}$ & $\mathbf{Y_5}$\\ \hline
        -0.0095361 & -0.0067635 & 0.0049676 & 0.0072610 & 0.0037596 \\ \hline
        $\mathbf{t_6}$ & $\mathbf{t_7}$ & $\mathbf{t_8}$ & $\mathbf{t_9}$ & $\mathbf{t_{10}}$ \\ \hline
        102.6433724 & 105.3788574 & 214.4145746 & 270.8216896 & 284.8259552 \\ \hline
        $\mathbf{Y_6}$ & $\mathbf{Y_7}$ & $\mathbf{Y_8}$ & $\mathbf{Y_9}$ & $\mathbf{Y_{10}}$ \\ \hline
        0.0055537 & -0.0095769 & 0.0012560 & 0.0075091 & -0.0054062 \\ \hline
       $\mathbf{t_{11}}$ & $\mathbf{t_{12}}$ & $\mathbf{t_{13}}$ & $\mathbf{t_{14}}$ & $\mathbf{t_{15}}$ \\ \hline
        306.1164378 & 321.7760341 & 321.7802714 & 371.5177862 & 397.2661914 \\ \hline
        $\mathbf{Y_{11}}$ & $\mathbf{Y_{12}}$ & $\mathbf{Y_{13}}$ & $\mathbf{Y_{14}}$ & $\mathbf{Y_{15}}$ \\ \hline
        0.0100000 & -0.0094967 & -0.0034873 & 0.0093145 & -0.0100000 \\ \hline
       $\mathbf{t_{16}}$ & $\mathbf{t_{17}}$ & $\mathbf{t_{18}}$ & $\mathbf{t_{19}}$ & $\mathbf{t_{20}}$  \\ \hline
        424.8279178 & 481.1300941 & 495.1351204 & 524.4073781 & 524.4973720 \\ \hline
        $\mathbf{Y_{16}}$ & $\mathbf{Y_{17}}$ & $\mathbf{Y_{18}}$ & $\mathbf{Y_{19}}$ & $\mathbf{Y_{20}}$ \\ \hline
        0.0042316 & -0.0041911 & 0.0045482 & 0.0084638 & -0.0098451 \\ \hline
        $\mathbf{t_{21}}$ & $\mathbf{t_{22}}$ & $\mathbf{t_{23}}$ & $\mathbf{t_{24}}$ & $\mathbf{t_{25}}$ \\ \hline
        548.5406082 & 564.6585589 & 564.6706719 & 670.4498730 & 677.6272855 \\ \hline
        $\mathbf{Y_{21}}$ & $\mathbf{Y_{22}}$ & $\mathbf{Y_{23}}$ & $\mathbf{Y_{24}}$ & $\mathbf{Y_{25}}$\\ \hline
        0.0100000 & -0.0100000 & -0.0021187 & -0.0080012 & 0.0047936 \\ \hline
       $\mathbf{t_{26}}$ & $\mathbf{t_{27}}$ & $\mathbf{t_{28}}$ & $\mathbf{t_{29}}$ & $\mathbf{t_{30}}$ \\ \hline
        695.0151092 & 704.9939567 & 709.7015756 & 714.4760969 & 714.9998480 \\ \hline
        $\mathbf{Y_{26}}$ & $\mathbf{Y_{27}}$ & $\mathbf{Y_{28}}$ & $\mathbf{Y_{29}}$ & $\mathbf{Y_{30}}$ \\ \hline
        -0.0060884 & 0.0029632 & 0.0009097 & -0.0100000 & -0.0100000 \\ \hline
        $\mathbf{t_{31}}$ & ~ & ~ & ~ & ~ \\ \hline
        715.0000000 & ~ & ~ & ~ & ~ \\ \hline
        $\mathbf{Y_{31}}$ & ~ & ~ & ~ & ~  \\ \hline
        0.0099010 & ~ & ~ & ~ & ~ \\ \hline
    \end{tabular}
    \caption{Patient LR01: Jump Times and Intensities of the Jumps , $N_Y$=31.}
    \label{saltietempiLR01}
\end{table}

\newpage

\begin{table}[!ht]
    \centering
    \begin{tabular}{||l|l|l|l|l||}
    \hline
        $\mathbf{t_1}$ &$\mathbf{t_2}$ & $\mathbf{t_3}$ & $\mathbf{t_4}$ & $\mathbf{t_5}$ \\ \hline
        0.0000000 & 0.9343427 & 1.5716458 & 6.4013948 & 7.1811615 \\ \hline
        $\mathbf{Y_1}$ & $\mathbf{Y_2}$ & $\mathbf{Y_3}$ & $\mathbf{Y_4}$ & $\mathbf{Y_5}$\\ \hline
        0.0047970 & -0.0092341 & 0.0070317 & -0.0074344 & 0.0027602 \\ \hline
       $\mathbf{t_6}$ & $\mathbf{t_7}$ & $\mathbf{t_8}$ & $\mathbf{t_9}$ & $\mathbf{t_{10}}$ \\ \hline
        15.5114491 & 16.0585447 & 21.0411272 & 47.7576120 & 60.6339624 \\ \hline
        $\mathbf{Y_6}$ & $\mathbf{Y_7}$ & $\mathbf{Y_8}$ & $\mathbf{Y_9}$ & $\mathbf{Y_{10}}$ \\ \hline
        0.0008573 & -0.0064724 & -0.0084104 & 0.0050180 & 0.0042248 \\ \hline
        $\mathbf{t_{11}}$ & $\mathbf{t_{12}}$ & $\mathbf{t_{13}}$ & $\mathbf{t_{14}}$ & $\mathbf{t_{15}}$ \\ \hline
        64.6421588 & 78.8809002 & 102.9121057 & 186.3817151 & 200.1306431 \\ \hline
       $\mathbf{Y_{11}}$ & $\mathbf{Y_{12}}$ & $\mathbf{Y_{13}}$ & $\mathbf{Y_{14}}$ & $\mathbf{Y_{15}}$ \\ \hline
        0.0084626 & 0.0050318 & -0.0100000 & -0.0027989 & 0.0019029 \\ \hline
       $\mathbf{t_{16}}$ & $\mathbf{t_{17}}$ & $\mathbf{t_{18}}$ & $\mathbf{t_{19}}$ & $\mathbf{t_{20}}$  \\ \hline
        286.6105518 & 349.5966979 & 360.9874552 & 437.1483624 & 466.2192794 \\ \hline
        $\mathbf{Y_{16}}$ & $\mathbf{Y_{17}}$ & $\mathbf{Y_{18}}$ & $\mathbf{Y_{19}}$ & $\mathbf{Y_{20}}$ \\ \hline
        0.0045002 & -0.0036133 & 0.0037398 & 0.0026732 & -0.0045150 \\ \hline
        $\mathbf{t_{21}}$ & $\mathbf{t_{22}}$ & $\mathbf{t_{23}}$ & $\mathbf{t_{24}}$ & $\mathbf{t_{25}}$ \\ \hline
        495.0856241 & 501.6031208 & 505.3289708 & 525.2767625 & 566.7834641 \\ \hline
        $\mathbf{Y_{21}}$ & $\mathbf{Y_{22}}$ & $\mathbf{Y_{23}}$ & $\mathbf{Y_{24}}$ & $\mathbf{Y_{25}}$\\ \hline
        -0.0016203 & 0.0034584 & 0.0024203 & -0.0068581 & 0.0034836 \\ \hline
         $\mathbf{t_{26}}$ & $\mathbf{t_{27}}$ & $\mathbf{t_{28}}$ & $\mathbf{t_{29}}$ & $\mathbf{t_{30}}$ \\ \hline
        590.0375421 & 590.4073183 & 615.1350938 & 620.5399033 & 661.3956095 \\ \hline
         $\mathbf{Y_{26}}$ & $\mathbf{Y_{27}}$ & $\mathbf{Y_{28}}$ & $\mathbf{Y_{29}}$ & $\mathbf{Y_{30}}$ \\ \hline
        0.0017887 & -0.0040903 & -0.0042061 & 0.0029953 & -0.0029417 \\ \hline
        $\mathbf{t_{31}}$ & $\mathbf{t_{32}}$ & $\mathbf{t_{33}}$ & $\mathbf{t_{34}}$ & $\mathbf{t_{35}}$ \\ \hline
        681.9175802 & 684.6887691 & 694.5082016 & 706.3630122 & 706.9024544 \\ \hline
        $\mathbf{Y_{31}}$ & $\mathbf{Y_{32}}$ & $\mathbf{Y_{33}}$ & $\mathbf{Y_{34}}$ & $\mathbf{Y_{35}}$ \\ \hline
        0.0080239 & -0.0098925 & -0.0099999 & 0.0064575 & -0.0086679 \\ \hline
        $\mathbf{t_{36}}$ & $\mathbf{t_{37}}$ & $\mathbf{t_{38}}$ & $\mathbf{t_{39}}$ &  \\ \hline
        711.9234096 & 714.7543400 & 714.8149993 & 714.9997922 & ~ \\ \hline
        $\mathbf{Y_{36}}$ & $\mathbf{Y_{37}}$ & $\mathbf{Y_{38}}$ & $\mathbf{Y_{39}}$ &  \\ \hline
        0.0050297 & -0.0099854 & -0.0094313 & 0.0099623 & \\ \hline
    \end{tabular}
     \caption{Patient MA01: Jump Times and Intensities of the Jumps , $N_Y$=39.}
    \label{saltietempiMA01}
\end{table}

\newpage

\begin{table}[!ht]
    \centering
    \begin{tabular}{||l|l|l|l|l||}
    \hline
       $\mathbf{t_1}$ &$\mathbf{t_2}$ & $\mathbf{t_3}$ & $\mathbf{t_4}$ & $\mathbf{t_5}$ \\ \hline
        2.0113565 & 2.3088969 & 4.1534739 & 10.1520456 & 10.2058897 \\ \hline
        $\mathbf{Y_1}$ & $\mathbf{Y_2}$ & $\mathbf{Y_3}$ & $\mathbf{Y_4}$ & $\mathbf{Y_5}$\\ \hline
        -0.0063838 & 0.0084358 & -0.0093116 & 0.0076411 & -0.0008623 \\ \hline
        $\mathbf{t_6}$ & $\mathbf{t_7}$ & $\mathbf{t_8}$ & $\mathbf{t_9}$ & $\mathbf{t_{10}}$ \\ \hline
        10.7173183 & 46.7399868 & 66.7684030 & 73.3555192 & 98.9393779 \\ \hline
        $\mathbf{Y_6}$ & $\mathbf{Y_7}$ & $\mathbf{Y_8}$ & $\mathbf{Y_9}$ & $\mathbf{Y_{10}}$ \\ \hline
        -0.0023263 & -0.0068971 & -0.0025987 & -0.0018733 & 0.0080688 \\ \hline
       $\mathbf{t_{11}}$ & $\mathbf{t_{12}}$ & $\mathbf{t_{13}}$ & $\mathbf{t_{14}}$ & $\mathbf{t_{15}}$ \\ \hline
        99.0424512 & 129.7995020 & 142.1732497 & 153.5438316 & 153.9491208 \\ \hline
        $\mathbf{Y_{11}}$ & $\mathbf{Y_{12}}$ & $\mathbf{Y_{13}}$ & $\mathbf{Y_{14}}$ & $\mathbf{Y_{15}}$ \\ \hline
        -0.0038734 & 0.0015776 & 0.0029459 & 0.0029387 & -0.0032109 \\ \hline
        $\mathbf{t_{16}}$ & $\mathbf{t_{17}}$ & $\mathbf{t_{18}}$ & $\mathbf{t_{19}}$ & $\mathbf{t_{20}}$  \\ \hline
        217.2682619 & 314.2622616 & 314.7854875 & 339.1542947 & 396.5938342 \\ \hline
        $\mathbf{Y_{16}}$ & $\mathbf{Y_{17}}$ & $\mathbf{Y_{18}}$ & $\mathbf{Y_{19}}$ & $\mathbf{Y_{20}}$ \\ \hline
        -0.0014188 & -0.0078609 & 0.0079105 & -0.0005204 & 0.0042779 \\ \hline
        $\mathbf{t_{21}}$ & $\mathbf{t_{22}}$ & $\mathbf{t_{23}}$ & $\mathbf{t_{24}}$ & $\mathbf{t_{25}}$ \\ \hline
        397.0592501 & 462.6756865 & 464.1050900 & 515.4325426 & 536.1261318 \\ \hline
       $\mathbf{Y_{21}}$ & $\mathbf{Y_{22}}$ & $\mathbf{Y_{23}}$ & $\mathbf{Y_{24}}$ & $\mathbf{Y_{25}}$\\ \hline
        -0.0063021 & 0.0082240 & -0.0093226 & 0.0028341 & -0.0039921 \\ \hline
        $\mathbf{t_{26}}$ & $\mathbf{t_{27}}$ & $\mathbf{t_{28}}$ & $\mathbf{t_{29}}$ & $\mathbf{t_{30}}$ \\ \hline
        557.9196158 & 578.4245305 & 578.5709511 & 614.1591499 & 615.0693980 \\ \hline
        $\mathbf{Y_{26}}$ & $\mathbf{Y_{27}}$ & $\mathbf{Y_{28}}$ & $\mathbf{Y_{29}}$ & $\mathbf{Y_{30}}$ \\ \hline
        0.0015155 & 0.0070801 & -0.0055078 & 0.0044835 & -0.0038320 \\ \hline
       $\mathbf{t_{31}}$ & $\mathbf{t_{32}}$ & $\mathbf{t_{33}}$ & $\mathbf{t_{34}}$ & $\mathbf{t_{35}}$ \\ \hline
        670.2118978 & 700.1335297 & 702.4794698 & 703.4086819 & 705.1864281 \\ \hline
        $\mathbf{Y_{31}}$ & $\mathbf{Y_{32}}$ & $\mathbf{Y_{33}}$ & $\mathbf{Y_{34}}$ & $\mathbf{Y_{35}}$ \\ \hline
        0.0026162 & -0.0012974 & 0.0084021 & -0.0097769 & 0.0046276 \\ \hline
        $\mathbf{t_{36}}$ & $\mathbf{t_{37}}$ & $\mathbf{t_{38}}$ & $\mathbf{t_{39}}$ & $\mathbf{t_{40}}$ \\ \hline
        712.2488239 & 712.4516812 & 714.1077618 & 714.7089113 & 714.9999888 \\ \hline
        $\mathbf{Y_{36}}$ & $\mathbf{Y_{37}}$ & $\mathbf{Y_{38}}$ & $\mathbf{Y_{39}}$ &  $\mathbf{Y_{40}}$\\ \hline
        0.0009019 & -0.0081884 & -0.0100000 & 0.0099993 & -0.0093725 \\ \hline
    \end{tabular}
     \caption{Patient PS01: Jump Times and Intensities of the Jumps , $N_Y$=40.}
    \label{saltietempiPS01}
\end{table}

\newpage

\begin{table}[!ht]
    \centering
    \begin{tabular}{||l|l|l|l|l||}
    \hline
        $\mathbf{t_1}$ &$\mathbf{t_2}$ & $\mathbf{t_3}$ & $\mathbf{t_4}$ & $\mathbf{t_5}$ \\ \hline
        0.0000000 & 0.0001144 & 0.0105762 & 3.2892763 & 3.5840276 \\ \hline
        $\mathbf{Y_1}$ & $\mathbf{Y_2}$ & $\mathbf{Y_3}$ & $\mathbf{Y_4}$ & $\mathbf{Y_5}$\\ \hline
        -0.0100000 & -0.0016208 & -0.0002454 & 0.0100000 & 0.0041683 \\ \hline
        $\mathbf{t_6}$ & $\mathbf{t_7}$ & $\mathbf{t_8}$ & $\mathbf{t_9}$ & $\mathbf{t_{10}}$ \\ \hline
        14.7586344 & 20.1801000 & 26.2047859 & 79.4419142 & 97.7874796 \\ \hline
        $\mathbf{Y_6}$ & $\mathbf{Y_7}$ & $\mathbf{Y_8}$ & $\mathbf{Y_9}$ & $\mathbf{Y_{10}}$ \\ \hline
        -0.0065450 & 0.0061877 & -0.0030495 & -0.0054302 & 0.0032026 \\ \hline
        $\mathbf{t_{11}}$ & $\mathbf{t_{12}}$ & $\mathbf{t_{13}}$ & $\mathbf{t_{14}}$ & $\mathbf{t_{15}}$ \\ \hline
        145.2549962 & 197.4705932 & 197.5688561 & 247.5807702 & 261.5608754 \\ \hline
       $\mathbf{Y_{11}}$ & $\mathbf{Y_{12}}$ & $\mathbf{Y_{13}}$ & $\mathbf{Y_{14}}$ & $\mathbf{Y_{15}}$ \\ \hline
        -0.0012758 & 0.0030801 & 0.0027528 & -0.0072852 & 0.0091927 \\ \hline
       $\mathbf{t_{16}}$ & $\mathbf{t_{17}}$ & $\mathbf{t_{18}}$ & $\mathbf{t_{19}}$ & $\mathbf{t_{20}}$  \\ \hline
        293.4605883 & 326.2834112 & 357.2401251 & 390.6000889 & 421.5138176 \\ \hline
        $\mathbf{Y_{16}}$ & $\mathbf{Y_{17}}$ & $\mathbf{Y_{18}}$ & $\mathbf{Y_{19}}$ & $\mathbf{Y_{20}}$ \\ \hline
        0.0039706 & -0.0100000 & 0.0088787 & -0.0053374 & 0.0099990 \\ \hline
       $\mathbf{t_{21}}$ & $\mathbf{t_{22}}$ & $\mathbf{t_{23}}$ & $\mathbf{t_{24}}$ & $\mathbf{t_{25}}$ \\ \hline
        423.9835398 & 465.8993583 & 471.8250029 & 475.8807161 & 488.3437979 \\ \hline
        $\mathbf{Y_{21}}$ & $\mathbf{Y_{22}}$ & $\mathbf{Y_{23}}$ & $\mathbf{Y_{24}}$ & $\mathbf{Y_{25}}$\\ \hline
        -0.0078473 & 0.0044494 & -0.0100000 & 0.0096852 & -0.0008359 \\ \hline
        $\mathbf{t_{26}}$ & $\mathbf{t_{27}}$ & $\mathbf{t_{28}}$ & $\mathbf{t_{29}}$ & $\mathbf{t_{30}}$ \\ \hline
        541.6525438 & 559.9605806 & 627.0197775 & 631.1670931 & 659.0129388 \\ \hline
       $\mathbf{Y_{26}}$ & $\mathbf{Y_{27}}$ & $\mathbf{Y_{28}}$ & $\mathbf{Y_{29}}$ & $\mathbf{Y_{30}}$ \\ \hline
        0.0073907 & -0.0100000 & 0.0100000 & -0.0089003 & -0.0093494 \\ \hline
       $\mathbf{t_{31}}$ & $\mathbf{t_{32}}$ & $\mathbf{t_{33}}$ & $\mathbf{t_{34}}$ & $\mathbf{t_{35}}$ \\ \hline
        669.0283123 & 688.2858345 & 691.7415747 & 706.6907198 & 712.1671379 \\ \hline
        $\mathbf{Y_{31}}$ & $\mathbf{Y_{32}}$ & $\mathbf{Y_{33}}$ & $\mathbf{Y_{34}}$ & $\mathbf{Y_{35}}$ \\ \hline
        0.0086358 & 0.0083106 & -0.0070506 & -0.0084924 & 0.0099757 \\ \hline
        $\mathbf{t_{36}}$ & $\mathbf{t_{37}}$ & $\mathbf{t_{38}}$ &  & \\ \hline
        712.8397024 & 714.9999999 & 715.0000000 & ~ & ~ \\ \hline
        $\mathbf{Y_{36}}$ & $\mathbf{Y_{37}}$ & $\mathbf{Y_{38}}$ &  & \\ \hline
        0.0008613 & 0.0100000 & 0.0099997 &  &\\ \hline
    \end{tabular}
    \caption{Patient TO01: Jump Times and Intensities of the Jumps , $N_Y$=38.}
    \label{saltietempiTO01}
\end{table}

%%%%%%%%%%%%%%%%%%%%%%%%%%%%%%%%%%%%%%%%%%%%%%
\section{Discussion}
\label{Discussion}

\noindent
In the present work a simple ODE model of night-time CGM data has been devised for each of eight experimental T2DM subjects with good metabolic control. It is clear that this work does not have any pretense of generality: rather, it endeavors to provide a proof of concept showing that the explicit representation of stochastic jumps allows the experimenter to interpret a noisy CGM signal without necessarily refer to a diet and exercise diary kept by the patient. Indeed, such diaries have been proven to be cumbersome to keep and of very limited value in the interpretation of the time course of glycemia, due to frequent omission of relevant events and due to the unavoidable imprecision in quantifying food intake and exercise levels. In addition, the (meta-)parameters estimated on the samples of waiting times and on the sample of jump intensities, together with such directly interpretable kinetic parameters such as $k_G$, may represent a first-line assessment of the status of the subject. These (meta-)parameters are easy to obtain repeatedly on non-invasive CGM tracings and could be useful in following a subject's clinical course. Routine, automated incorporation of the model within glycemia-tracking software apps may help to continuously monitor the state of health of a patient at no extra cost and with no added invasiveness.

%%%%%%%%%%%%%%%%%%%%%%%%%%%%%%%%%%%%%%%%%%%%%%%%%%%%%%%%%



\section*{Acknowledgements}
\noindent
The authors Giulia Elena Aliffi, Giovanni Nastasi and Vittorio Romano acknowledge the support from INdAM (GNFM).

\section*{Declaration of Interests}
\noindent
None

\section*{Fundings}
\noindent
This research did not receive any specific grant from funding agencies in the public, commercial, or not-for-profit sectors.

\section*{Credit authorship contribution statement}

\noindent
\textbf{Giulia Elena Aliffi:} Conceptualization, Methodology, Software, Formal Analysis, Investigation, Writing-original draft, Writing-review \& editing.
\textbf{Giovanni Nastasi:} Methodology, Writing-review \& editing.
\textbf{Vittorio Romano:} Methodology, Formal Analysis, Writing-review \& editing, Supervision, Project Administration.
\textbf{Dario Pitocco:} Resources, Data curation, Writing-review \& editing.
\textbf{Alessandro Rizzi:} Resources, Data curation, Writing-review \& editing.
\textbf{Elvin J. Moore:} Methodology, Formal Analysis, Writing-review \& editing.
\textbf{Andrea De Gaetano:} Conceptualization, Methodology, Writing-review \& editing, Supervision.\\

\noindent
All authors have read and agreed to the published version of the manuscript.

\end{document}